\documentclass[iop,revtex4-1]{emulateapj}
\usepackage{natbib}
\usepackage{hyperref}
\usepackage{newtxtext,newtxmath}
\usepackage{scalerel}
\usepackage{amsmath}	% Advanced maths commands
\usepackage{amssymb}	% Extra maths symbols
\usepackage[dvipsnames]{xcolor}

\newcommand{\mytilde}{\raise.19ex\hbox{$\scriptstyle\sim$}}
\renewcommand{\vec}[1]{\boldsymbol{#1}}
\begin{document}

\title{RECONSTRUCTION OF RADIO RELICS AND X-RAY TAILS IN AN OFF-AXIS CLUSTER MERGER: HYDRODYNAMICAL SIMULATIONS OF A115}

\shorttitle{Merger Scenario of A115}
\shortauthors{Lee et al.}
%% Note that the corresponding author command and emails has to come
%% before everything else. Also place all the emails in the \email
%% command instead of using multiple \email calls.

\author{Wonki Lee\altaffilmark{1}, M. James Jee\altaffilmark{1,2}, Hyesung Kang\altaffilmark{3},
Dongsu Ryu\altaffilmark{4},  Taysun Kimm\altaffilmark{1}, and Marcus Br\"uggen\altaffilmark{5}
}
\altaffiltext{1}{Department of Astronomy, Yonsei University, 50 Yonsei-ro, Seoul 03722, Korea; wonki.lee@yonsei.ac.kr, mkjee@yonsei.ac.kr}
\altaffiltext{2}{Department of Physics, University of California, Davis, One Shields Avenue, Davis, CA 95616, USA}
\altaffiltext{3}{Department of Earth Sciences, Pusan National University, Busan 46241, Korea}
\altaffiltext{4}{Department of Physics, School of Natural Sciences, UNIST, Ulsan 44919, Korea}
\altaffiltext{5}{Hamburger Sternwarte, Universit\"at Hamburg, Gojenbergsweg 112, 21029 Hamburg, Germany}

\begin{abstract}
    Although a merging galaxy cluster is a useful laboratory to study many important astrophysical processes and fundamental physics, only limited interpretations are possible without careful analysis of the merger scenario. However, the study is demanding because a thorough comparison of multi-wavelength observations with detailed numerical simulations is required.
    In this paper, we present such a study for the  off-axis binary merger A115. The system possesses a number of remarkable observational features, but no convincing merger scenario, explaining the shape and location of the radio relic in harmony with the orientation of the cometary X-ray tails, has been presented.
    Our hydrodynamical simulation, with adaptive mesh refinement, suggests that the cometary X-ray tail of A115 might be a slingshot tail and can arise $\mytilde0.3$~Gyrs after the impact and before the two subclusters reach their apocenters. This scenario can predict the location and orientation of the giant radio relic, which is parallel to the northern X-ray tail. 
    In addition, our study indicates that diffusive shock acceleration alone cannot generate the observed radio power unless aided by  stronger magnetic fields and/or more significant presence of fossil electrons.
\end{abstract}


\section{Introduction}
\label{sec:intro}
Collisions of galaxy clusters are among the most energetic astrophysical processes in the universe.
An immense amount of energy on the order of $\mytilde 10^{64} \, \rm erg$ \citep[][]{Ricker2001} is transferred to the cluster galaxies and their environments, which results in excitation of the motion of the cluster galaxies \citep{Pinkney1996,Takizawa2010}, heating and turbulence of the intracluster medium \citep[ICM,][]{Ricker2001,Ryu2008,Vazza2012}, and creation of merger shocks propagating up to a few Mpcs \citep{Ensslin1998,Bonafede2014,VanWeeren2019}. 
Therefore, merging clusters are useful laboratories to test our understanding of a number of important astrophysical processes. More recently, they are also considered cosmic particle accelerators, where fundamental properties of dark matter can be studied~\citep[e.g.,][]{randall2008b,Harvey2015,Wittman2018b}.

However, several challenges prevent us from fully utilizing their profound potentials as invaluable laboratories. One outstanding challenge is the merger phase ambiguity \citep{Wittman2018a}. Based on limited data, one cannot assign a unique merger scenario to the observed system. For example, 
an observed X-ray tail may originate from either ram pressure stripping or sloshing; the latter requires a different merger phase than the former \citep[e.g.,][]{Lyskova2019}. Although extensive, multi-wavelength data help to break the merger scenario degeneracy. In particular, clusters with so-called ``radio relics" are sure signs of post-mergers because the radio features are believed to arise from acceleration of cosmic ray particles across the merger shock. This high signal-to-noise shock-tracing capability is one of the exclusive merits of radio observations, not easily accessible by X-ray observations. The combination of spectroscopic observations of the cluster galaxies and polarization data of the radio relics enables us to constrain the viewing angle. Weak lensing provides unbiased masses of individual substructures \citep[e.g.,][]{Jee2016,Finner2017}, which are expected to depart from hydrostatic equilibrium.

One of the key components in merger scenario reconstruction is a high-fidelity hydrodynamical simulation. Although multi-wavelength data alone can provide rough, qualitative constraints, it is through iterative simulations that quantitative analysis is achieved. A number of merger simulations have focused on ICM properties and been quite successful in reproducing some of observed features in X-ray observations \citep[e.g.][]{Lage2014}. However, there have been much fewer studies that address observed properties of radio relics. This lack of numerical studies, attempting to reproduce observed properties of radio relics, is understandable because only recently have some state-of-the-art codes have begun to handle particle acceleration by injecting cosmic rays as a fluid using the on-the-fly shock detection scheme \citep[e.g.][]{Dubois2016}. 
In addition, little is known as to the creation and evolution of magnetic fields in the ICM.
Nevertheless, through post-processing of hydrodynamical simulation data, it is possible to identify shocks and estimate synchrotron radio emissivities. Although this requires several assumptions on the details of the astrophysical process, some authors have demonstrated that the method reasonably reproduces the locations and orientations of radio relics in merging clusters \citep[e.g.,][]{Bruggen2012,Hong2015,Vazza2016}.

\begin{figure*}
\centering
\includegraphics[width=2\columnwidth]{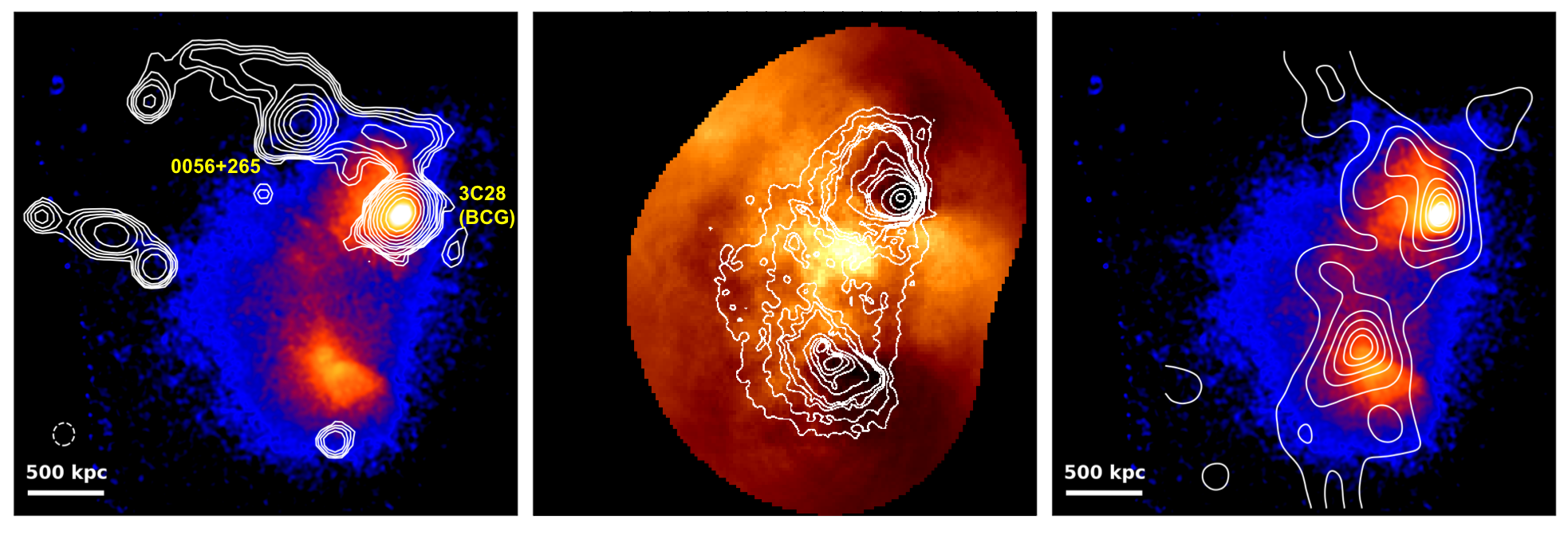}
\caption{Observations of A115 from various sources. (Left) The color is the \textit{Chandra} X-ray image obtained after the combination of all archival data (361 ks). Point sources are removed and exposures are corrected for using the {\tt CIAO}~4.11 package. 
White contours show the VLA D array L band (1.4 GHz) observation. The VLA data are processed with the task {\tt vlarun} within {\tt AIPS}. The dashed circle indicates the beam size ($42\arcsec\times 40\arcsec$). The contour levels are $3\sigma  \times \sqrt{2}^n$, where $1\sigma =0.25 \rm \, mJy~beam^{-1}$ and $n=0,1,2,3,4,6,8,10,12,14,16,18$.
The eastern radio structure is a tailed radio galaxy \citep{Govoni2001}. 
We denote the location of the two known point radio sources 0056+265 and 3C28.
The orientation of the cometary tail of A115N is approximately parallel to that of the radio relic, whose western side touches the northern edge of A115N.
(Middle) The \protect\cite{Hallman2018} temperature map is shown. The value ranges from $4\,\rm keV$ (black) to $16 \rm\,keV$ (yellow). The white contour represents the X-ray flux density.
(Right) The \protect\cite{Kim2019} weak lensing mass contours are overlaid on the \textit{Chandra} X-ray image. Note the good alignments of the peaks between X-ray and mass.
\label{fig1}}
\end{figure*}

In this paper, we present our hydrodynamical simulations of Abell~115 (hereafter A115) with the {\tt RAMSES} adaptive mesh refinement (AMR) code \citep{Teyssier2002}. 
This is the first study that simulates both X-ray and radio properties of the system with an idealized simulation setup.
A115 is an off-axis binary cluster merger confirmed with the X-ray emission \citep[e.g.][]{Forman1981}, the member galaxies \citep[e.g.][]{Barrena2007}, and the WL mass distributions \citep[e.g.][]{Kim2019}.
The X-ray data show two distinct cool cores. Both the northern and southern subclusters (hereafter A115N and A115S, respectively)
possess a characteristic tail-like morphology (Figure~\ref{fig1}), which has been attributed to ram-pressure stripping \citep[e.g.,][]{Gutierrez2005}. Some authors often further interpret this peculiar morphology as indicating that the two subclusters might be rotating around the center of mass.
%\textcolor{red}{However, a giant ($\mytilde2 \rm \,Mpc$) radio relic oriented nearly parallel to the velocity vector of A115N indicated by the X-ray tail direction has been observed \citep{Govoni2001}. In addition, \cite{Botteon2016} report the rare case of coexisting X-ray shock over the radio relic and further support the presence of the merger shock, yet in a puzzling orientation because the inferred motion of the subcluster from this merger shock is perpendicular to the velocity vector from the X-ray tail~\citep[e.g.,][]{Golovich2018}.}
However, the presence of the giant ($\mytilde2 \rm \,Mpc$) radio relic coincident with the X-ray shock features \citep{Botteon2016} is somewhat puzzling because the motion of the subcluster implied by the orientation of the relic is perpendicular to the velocity vector inferred by the X-ray tails~\citep[e.g.,][]{Golovich2018}.
In an attempt to resolve the puzzle, \cite{Hallman2018} suggest a scenario, wherein the observed giant radio relic is part of the early bow shock formed as A115N moves in the direction indicated by the tail (while orbiting around A115S).
Although this is an interesting possibility, their merger scenario reconstruction is based on the analysis of the A115 analogs in cosmological hydrodynamical simulations, where the resolution for the bow shock is low and the simulated cluster properties do not match A115. 

Therefore, in order to enable more quantitative analysis, we perform idealized simulations of A115, utilizing our multi-wavelength observation results to set up simulation configurations.
We do not attempt to fine-tune the simulation setups exhaustively to reproduce all observed features. Instead, we focus on reproduction of the morphology, orientation, and location of both X-ray emission and radio relic, which is one of the main outstanding puzzles in A115.
Since there has been no idealized hydrodynamical simulation study with the same goal, the result from the current study will also facilitate the interpretation of other off-axis cluster mergers with radio relics.

This paper is structured as follows. We start with the review of the previous multi-wavelength studies of A115 in \textsection\ref{sec:a115}.
The hydrodynamical simulation setups and procedures needed to generate mock-observations are described in \textsection\ref{sec:num}.
We present our results and discussions in \textsection\ref{sec:res} and \textsection\ref{sec:dis}, respectively, before we summarize the paper in \textsection\ref{sec:con}.

\section{Summary of Observational Features in A115}
\label{sec:a115}
A115 is a galaxy cluster at $z=0.197$ with two 
  X-ray cores separated by $\mytilde900\,\rm kpc$.
  Both cores possess cometary tails stretched in opposite directions (see left panel of Figure \ref{fig1}).
  The deep \textit{Chandra} observations have revealed that the intermediate region between the cores is very hot ($T \sim11 \, \rm keV$), while each core is cool
  \citep[\(T_{\rm A115N} \sim 3\, \rm keV \, \mathrm{and} \, T_{\rm A115S} \sim 4\, \rm keV\),][see middle panel of Figure \ref{fig1}]{Hallman2018}.
  The cometary tails and the hot region indicate that the system is undergoing a major merger whereas the intactness of the cool cores is attributed to an off-axis merger with a non-negligible impact parameter ~\citep[e.g.][]{Golovich2018}.
  
VLA observations (Project ID: AF349) have uncovered a giant ($\mytilde 2 \, \rm Mpc$) radio relic with a few  compact radio sources associated with the cluster redshift (see Figure~\ref{fig1} for the location of the two radio sources 3C28 and 0056+265).
  The Mach number of the relic is estimated to be $\mathcal{M}_{\rm DSA} \sim 4.6$ from the spectral slope $\alpha \sim 1.1$ \citep{Govoni2001}.
  
  Very few galaxy clusters have been reported to show shock features traced by both radio and X-ray observations simultaneously. A115 is one of these rare examples. \cite{Botteon2016} detected both density and temperature jumps across the radio relic using the {\it Chandra} data, showing that the Mach numbers derived from these two measurements are consistent with each other.\footnote{However, \cite{Hallman2018} report a pressure jump near the western edge of the A115N X-ray core, not the northern edge where the radio relic is located.} The density (temperature) jump gives $\mathcal{M}=1.7\pm 0.1$ ($1.8_{-0.4}^{+0.5}$). These values are a factor of two lower than the radio relic result, which assumes diffusive shock acceleration (DSA) \citep[\(\mathcal{M}_{\rm DSA} \sim 4.58^{+\infty}_{-2.50} \),][]{Govoni2001}.
  
  The coincidence of the radio diffuse emission with the X-ray shock features supports the possibility of the emission being indeed a relic.
  However, as noted by \cite{Govoni2001}, in general, 
  radio relics are observed in the periphery of the system. Thus, the proximity of the A115 radio emission to A115N is somewhat unusual, given the large separation between A115N and A115S. Exacerbating the puzzle, the orientation of the radio relic is almost parallel to the X-ray tail of A115N. If we interpret the X-ray tail as indicating the direction of the motion, the orientation of the current radio relic is perpendicular to our prediction.
  
  A115 was recently revisited with Subaru/SuprimeCam and Keck/DEIMOS \citep{Kim2019} as part of the project Merging Cluster Collaboration \citep[e.g.][$MC^2$]{Golovich2016}.
  The \cite{Kim2019} WL analysis reveals two mass clumps in good spatial agreement with the cluster galaxies and X-ray peaks (right panel of Figure \ref{fig1}).
  The total mass of the A115 system is $M_{200} = 6.41^{+1.08}_{-1.04} \times 10^{14} \,M_{\odot}$ whereas the individual A115S and A115N masses are $M_{200} = 3.15^{+0.79}_{-0.71} \times 10^{14} \,M_{\odot}$ and $ 1.58^{+0.56}_{-0.49} \times 10^{14} \,M_{\odot}$, respectively \citep{Kim2019}.
  These WL masses are at maximum an order of magnitude lower than dynamical masses based on $\mytilde250$ spectroscopic redshifts \citep{Golovich2018}, indicating that the on-going merger might have inflated the velocity dispersion measurements significantly.
  
  \cite{Golovich2018} and \cite{Kim2019} claim that
  the collision is happening nearly in the plane of the sky ($\theta\sim25 ^{\circ}$) based on the small line-of-sight (LOS) velocity difference $V_{\rm LOS} \sim 230\pm100~ \rm km ~\rm s^{-1}$ between A115N and A115S.
  Interestingly, if the velocity analysis is confined to the cluster members within 0.25 Mpc of the brightest cluster galaxies (BCGs), the LOS velocity difference increases to $838\pm549~ \rm km~s^{-1}$.
  However, this is based on small numbers (15 members for each subcluster) and thus lacks statistical significance \citep{Kim2019}.
    
In summary, the panchromatic view on A115 has hinted that A115 is a result of an off-axis major merger occurred in the plane of the sky.
However, the peculiar configuration of the radio relic and X-ray morphology is still to be explained with a coherent merger scenario, which is the primary goal of the current study.
\section{Numerical Simulations of A115 and 
Conversion to Mock Observations}
\label{sec:num}

\subsection{Simulation Setups}
 To understand the merger history of A115 that is consistent with the observed features in radio and X-ray, we perform  idealized hydrodynamics simulations of off-axis cluster mergers using {\tt RAMSES} \citep{Teyssier2002}.
 The Euler equations for hydrodynamics are solved using the HLLC scheme \citep{Toro1994} with a Courant number of 0.5 and Min-Mod limiter in the slope computation.
 
   %Main, sub
   We design our collision simulation in an isolated box with a volume of $15^3 \, \mathrm{Mpc^3}$ using outflow (zero-gradient) boundary conditions. %15^3
   The cells are resolved based on their density from the lowest level (cell size of $0.46 \rm \, Mpc$) to the highest level of refinement (cell size of $7.32 \rm \, kpc$).
   With higher resolution simulations ($\mytilde3.6 \rm \, kpc$), we verify that this resolution gives converging features that we present in this paper.
   An additional refinement criterion based on the pressure gradient \citep[\(\Delta P / P>1 \),][]{Teyssier2002} is applied so that the shock structures propagating at the cluster periphery are also maximally resolved.
   This allows us to model merger shocks outside the virial radius of both clusters with $\mytilde$20 million leaf cells at the maximum refinement level, out of the total 30 million leaf cells.
   To reduce complexity, neither radiative cooling nor baryonic feedback is implemented.
   We believe that the exact morphology of the X-ray surface brightness distribution depends on the implementation details and thus should be a subject of future studies. Nevertheless,
   \cite{Kang2007} and \cite{Hong2015} show that these two processes have only minor impacts on shock properties that we focus on in the current paper.

   Two clusters, the more massive main cluster (hereafter with subscript `m') and the less massive sub cluster (subscript `s'), are generated to represent A115S and A115N at the cluster redshift, respectively.
   Both clusters are spherically symmetric, comprised of dark matter and the ICM.
   We let dark matter follow an Navarro-Frenk-White \citep[NFW,][]{Navarro1996} profile and the ICM a beta profile \citep{Cavaliere1976}. The two profiles are:
   \begin{equation}
      \rho_{\scaleto{\rm DM}{4pt}} (r) = \frac{\rho_{0,\rm DM}}{r/r_{\rm s}(1+r/r_{\rm s})^2}
      \label{eq:DMprof}
    \end{equation}
    \vspace{-7pt}
    \begin{equation}
        \rho_{\scaleto{\rm ICM}{4pt}} (r)=\frac{\rho_{0,\rm ICM}}{(1+r^2/r_{\rm c}^2)^{\frac{3}{2}\beta}}
  \label{eq:ICMprof}
  \end{equation}
  \noindent
  where  $\rho_{0,\rm DM}$ and $\rho_{0,\rm ICM}$ are the normalization constants, $r_{\rm s} = r_{200}/c$ is the scale radius of the dark matter halo, and $r_{\rm c}$ is the ICM core radius.
  We use $\beta = 2/3$ (i.e. isothermal profile) in our simulations.
  We adopt the scale radius $r_{\rm s}$ and total mass $M_{200}$ of dark matter halo from
  the latest WL study of \cite{Kim2019}. 
  Then we determine the ICM density normalization $\rho_{0,\rm ICM}$ so that it satisfies the baryon mass fraction $f_{\rm{bar},200} \sim 0.13$ at $r_{200}$.
  We choose $r_{\rm c} = 40 \rm \,kpc$ ($30\,\rm kpc$) for the main (sub) cluster. Both
  ICM core radii are small fractions (10\% --16\%) of the dark matter halo scale radii $r_s$ and are motivated by the presence of the cool cores. 
  Using the Disk Initial Conditions Environment \citep[DICE;][]{Perret2016} code, we generate $10^7$ dark matter particles for each cluster and assign gas densities to the individual AMR cells.
  The temperature for each cell is computed in such a way that an hydrostatic equilibrium is satisfied at the given potential.

  Since both equations~\ref{eq:DMprof} and~\ref{eq:ICMprof} give a diverging total mass, we make the density drop exponentially beyond
  the cut-off radii. Following  the suggestion of \cite{Donnert2017}, we choose $r_{\rm cut}=2.2 \,\rm Mpc$ and $1.8  \,\rm Mpc $ for the main and sub clusters, respectively, for both DM and ICM profiles.
  The exponential decrease of the gas density stops when it reaches the background density $\rho_{\rm back} = 1.5 \times 10^{-30} \rm \,g/cm^3$. We assign $T_{\rm back} = 10^6 \rm K$ to the background temperature, which is also the minimum ICM temperature of each cluster. This is needed to achieve the pressure equilibrium  at the boundary between the cluster and background.
  
  We tested the dynamical stability of each halo by simulating it in isolation for $4\,\rm Gyr$.
  We use the change of the two radii that enclose 50\% of the DM and the ICM, respectively, as a measure of the stability. We find that both radii decrease by up to 10\% during the first $\mytilde 2\rm\, Gyr$, but afterwards they remain constant.
  Because our collision is designed to take place at $\mytilde4\rm\, Gyr$ after the initial separation, our simulated clusters are in the stable state at the impact.

  Finally, the two clusters are positioned on the $x-y$ plane with a  separation of $\Delta y=4.3\rm Mpc$.
  The initial velocities are chosen to produce a collision velocity of $\mytilde 2,000 ~ \rm km~ s^{-1}$ at the impact.
  Also, we apply an initial offset in the $x$ direction to obtain a desired pericenter distance, $b\sim500~\rm kpc$.
  The collision velocity $\mytilde 2,000~\rm km~ s^{-1}$ is slightly greater than the free-fall velocity $\mytilde1,800~\rm km~ s^{-1}$ estimated with the timing argument \citep{Sarazin2002}. In Appendix~\ref{app1}, we show that this collision velocity is bracketed by the range of the values observed in the cosmological simulation.
  The initial condition parameters are summarized in Table \ref{tab:ref} and \ref{tab:orbit} and we refer to this setup as the reference run.
  %Both the collision velocity and pericenter distance for this reference are chosen so that the results reasonably reproduce the observed high Mach number ($\mathcal{M}_{\rm DSA} \sim 4.58$) with sustained cool cores, although, as mentioned earlier, we do not exhaustively search for the optimal combination.
  Both the initial velocity and pericenter distance for this reference are chosen so that the results reasonably reproduce the observed high Mach number ($\mathcal{M}_{\rm DSA} \sim 4.58$) with sustained cool cores.
  
 As mentioned in \textsection\ref{sec:intro}, the current paper focuses on the reproduction of the morphology, orientation, and location of both the X-ray emission and radio relic of A115. These results depend on parameters such as the collision velocity, pericenter distance, and viewing angle; for instance, a lower initial velocity results in a more prominent X-ray tail because the X-ray gas suffers from a higher ram pressure.
  Therefore, in order to investigate their effects, we run three additional simulations by modifying the initial velocity and impact parameter from the reference run.
  These include two simulations with modified velocities in the $y$ direction by $\pm200 \,\rm km~ s^{-1}$ and one simulation with a larger ($\Delta b=+600 \rm \,kpc$) impact parameter, which in turn increases the pericenter distance by $300 \rm \,kpc$. Hereafter, we refer to these three setups as the $V+$, $V-$, and $b+$ runs, respectively. We also investigate one case when the merger axis is not exactly perpendicular to the line-of-sight direction. We are aware that different combinations of various initial velocities, impact parameters, and viewing angles can lead to non-negligible differences in the final results. However, it is infeasible to exhaustively search for the optimal combination in this  computationally expensive AMR study whose effective dynamic range in volume is $2048^3$.

\begin{table}
	\centering
	\caption{Halo properties in the simulation. }
	\label{tab:ref}
	\begin{tabular}{ccc} % four columns, alignment for each
		\hline
		  & Main & Sub\\
		\hline
	    $\mathrm{M}_{200}[10^{14} M_{\odot}]$ & 3.15 & 1.58\\
		$R_{\rm s}$ [kpc]&   380 & 280\\
		$R_{\rm c}$ [kpc] &   40  & 30 \\
		%$V_{y}$ [km/s] & -100 & 600 \\
        \hline
	\end{tabular}
\end{table}
\begin{table}
	\centering
	\caption{Orbital properties for the reference run.}
	\label{tab:orbit}
	\begin{tabular}{cc} % four columns, alignment for each
		\hline
		 Property & Value\\
		\hline
	   initial relative velocity  & 700 km/s\\
	   impact velocity   & 2000 km/s\\
	   pericenter distance  & 500 kpc \\
        \hline
	\end{tabular}
\end{table}  

\subsection{Generation of Mock Observation}
\subsubsection{X-ray Mock Observation}
  The X-ray emissivity of each gas cell is calculated based on the following description of bremsstrahlung radiation from thermal electrons in a fully ionized gas with a primordial composition \citep{Marinacci2018}:
 
\begin{equation}\label{eq:Xray}
\begin{split}
      \epsilon_{\scaleto{\mathrm{X-ray}}{5pt}} \,  =\, &4.90 \times 10^{-28}  \rm \, erg \: s^{-1} \\ 
      & \times n_{\rm H} ^2 T^{0.5}\left[\mathrm{exp}\left(-\frac{T_1}{k_{\rm B} T}\right)-\mathrm{exp}\left(-\frac{T_2}{k_{\rm B} T}\right)\right],
\end{split}
\end{equation}

  \noindent
  where $n_{\rm H}$ and $T$ are the hydrogen number density (in units of $\rm cm^{-3}$) and temperature of each cell (in units of $\rm K$), respectively. Note that we do not include metal lines, which can non-negligibly alter the results in the low-temperature region, but are not likely to significantly modify the results in the high-temperature region that we focus on in the current study.
    We choose $T_1 = 0.5\rm \,keV$ and $T_2 = 7\rm \,keV$ in order to match the energy range used in our \textit{Chandra} image creation.
  The X-ray surface brightness map is obtained by integrating the emissivity contributed from every cell along the line-of-sight ($z-$axis) direction.

  The {\it Chandra} temperature map shown in the middle panel of Figure \ref{fig1} \citep[][]{Hallman2018} is estimated from the local X-ray spectral fitting.
  To simulate this {\it Chandra} spectroscopic temperature map, we average the temperature of each cell using the weight suggested by \cite{Mazzotta2004}:
  \begin{equation}\label{eq:Temp}
      T_{\rm spec} = \frac{\sum_{\scaleto{\mathrm{LOS}}{4pt}} n_i^2 T_i^{0.25} dx^3_i}{\sum_{\scaleto{\mathrm{LOS}}{4pt}} n_i^2 T_i^{-0.75} dx^3_i},
  \end{equation}
  \noindent
  where $n_i, T_i$, and $dx_i$ represent the gas density, temperature, and cell size of the $\mathrm{i^{th}}$ cell along the line-of-sight ($z-$axis) direction.
  
\begin{figure*}
    \centering
	\includegraphics[width=2\columnwidth]{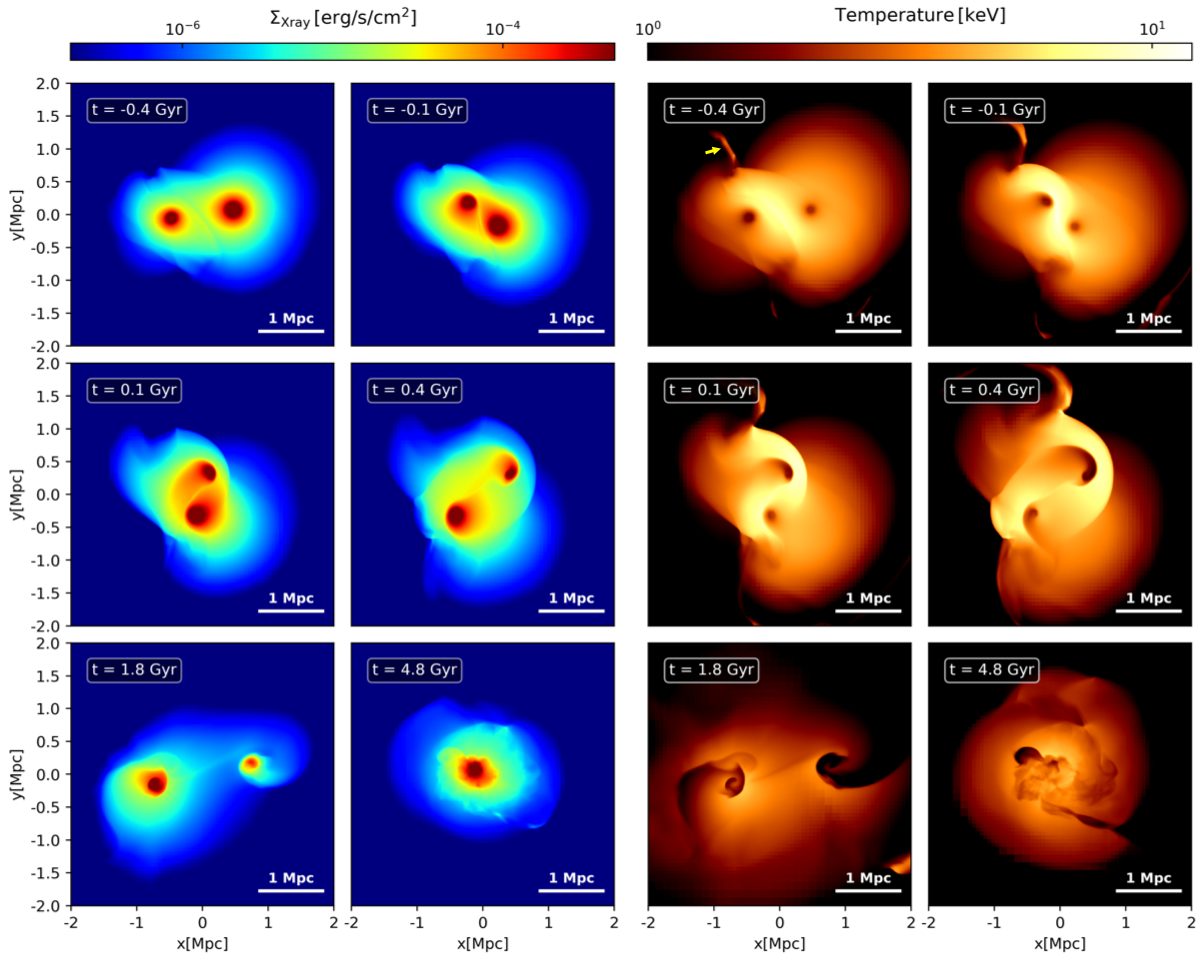}
    \caption{Simulated X-ray surface brightness (first and second columns) and temperature
    (third and fourth columns) maps. We display the results from the reference run at six different epochs. The $t=0$~Gyr snapshot corresponds to the epoch, when the two clusters are at the smallest separation.
    These maps are generated using the equations described in \textsection\ref{sec:num}. 
    The compression of the gas is observable as early as $t=-0.4\rm \,Gyr$. At $t=-0.1 \rm \,Gyr$ the shocks become distinct and together with the compressed gas form the ``S" shape in the temperation map. Soon after the closest passage ($t=0.1 \rm \,Gyr$), we can witness the long ram-pressure tails, whose directions are approximately parallel to the velocities of the cluster cores. At $t=0.4 \rm \,Gyr$, the ram-pressure tails disappear whereas the slingshot tails are emerging.
    The on-set of the second in-fall happens at $t=1.8 \rm \,Gyr$.  The $t=4.8 \rm \,Gyr$
    epoch is the third passage, when we observe turbulent structures. The yellow arrow in the temperature map at $t=-0.4\rm\,Gyr$ points to the ``temperature fold" feature (see text).}
    \label{fig2}
\end{figure*}
  
  \subsubsection{Shock detection algorithm}
  \label{sec:shock}
  We adopt the shock criteria suggested by \cite{Ryu2003} and follow the implementation for the AMR structure described by \cite{Skillman2008}.
  We provide a brief summary below.
  
  A cell is flagged as a shock center candidate if
  its gas velocity converges ($\nabla \cdot \vec{\rm v}<0$).
  We walk through the candidate cells in the $\pm\nabla T$ directions and find the cell that first violates either of the two conditions: $ \Delta T \times \Delta S >0$ or $\mathrm{log}\left|\Delta T\right| >0.11$, where  $\Delta T$ and $\Delta S$ are the temperature and entropy differences between the two adjacent cells. 
  The cell that first violates either of the two conditions is classified as a post-shock (pre-shock) cell if the search direction is $\nabla T>0$ $(<0)$.
  Prior to identifying both pre-shock and post-shock cells, if the divergence,  $\nabla \cdot \vec{\rm v}$, of the visited cell is found to be lower than the shock center candidate that initiated this walk, the shock center candidate is discarded.
  Once both post- and pre-shock cells are identified, the shock center candidate is promoted to a shock center with a Mach number ($\mathcal{M}$) derived from the following relation: 
  \begin{equation}
      T_2/T_1 = \frac{(5\mathcal{M}^2-1)(\mathcal{M}^2+3)}{16\mathcal{M}^2},
  \end{equation}
  where $T_1$ and $T_2$ are the temperatures of the pre- and the post-shock cells, respectively.
  Thus, each shock structure is a column of cells comprised of one shock center, one post- and one pre-shock cells, and multiple shock cells [see Figure 1 of \cite{Skillman2008} for the schematic description].
  Note that we must perform this shock identification with the smallest AMR cells in order to avoid facing multiple adjacent cells. However, we verify that no cells at a coarser level satisfy our shock conditions described above. This is attributed to our refinement criteria $\Delta P / P>1 $, which ensures that the shock cells are always maximally resolved.%the smallest cells.
  
  \subsubsection{Generation of Mock Radio Observation}
  \label{sec:mockradio}
  The radio emissivity is estimated using the prescription of \cite{Hong2015}.
  For each shock structure, we first compute the kinetic energy flux $f_{\rm k}$, the main power source of the radio emission, as follows: 
  \begin{equation}
     f_{\rm k} = \frac{1}{2} \rho (\mathcal{M}c_{\rm s})^3,
     \label{eq:fk}
  \end{equation}
  \noindent
  where $c_{\rm s}$ is the sound speed of the pre-shock cell given by $c_s = (\gamma P/\rho)^{1/2}$ with $\gamma = 5/3$.
  Then, we calculate the efficiency of acceleration as an increasing function of the Mach number $\mathcal{M}$.
 \cite{Kang2013} show that the dissipation efficiency $\eta(\mathcal{M})$ becomes negligible for $\mathcal{M}<3$, approaches $\mytilde0.8\%$ at $\mathcal{M}=3$, and saturates to $\mytilde20\%$ at $\mathcal{M}>10$. 
 This determines the momentum distribution of cosmic ray electrons (CRe) for each shock structure:
  \begin{equation}
     f_{\rm CRe}(p) \propto K_{\rm e/p} \eta(\mathcal{M}) f_{\rm k} p^{-\rm q},
  \end{equation}
  \noindent
  where $q=3\sigma /(\sigma -1)$ is the spectral index, 
  $\sigma$ is the compression rate given by
   $\sigma=(\gamma+1)\mathcal{M}^2/[(\gamma-1)\mathcal{M}^2+2]$, $p$ is 
  the momentum of electron, and $K_{\rm e/p}$ is the CR electron-to-proton ratio, which is set to $0.01$ \citep{Hong2015}.
  Following \cite{Hong2015}, we also apply the exponential cut-off due to cooling in the momentum distribution.
  Finally, the radio emissivity ($j_{\nu}$) is obtained from
   \begin{equation}
     j_{\nu} \propto \int_{p_{\rm min}}^{\infty} B F\left(\frac{\nu}{\nu_{\rm c}(p)}\right) f_{\rm CRe} d^3p,
     \label{j_nu}
  \end{equation}
  \noindent
  where $B$ is the magnetic field strength of the post-shock cell,
  $\nu_{\rm c}$ is the characteristic frequency, and 
  $F(x) \equiv x\int_{x}^{\infty}d\eta K_{5/3}(\eta)$ with $K_{5/3}(\eta)$ being a modified Bessel function.
  In our study, we assume a constant magnetic field of $B=1\,\mu G$, which is derived by \cite{Govoni2001} from the equi-partition assumption at the A115 radio relic.
 The minimum momentum $p_{\rm min}$ contributing to the radio emission  is set to $p_{\rm min} = 0.01m_{\rm p}c$, which assumes that DSA acts on $p\gtrsim3p_{\rm th,p}$, where $p_{\rm th,p}$ is the most probable momentum of thermal protons.
  Accordingly, radio emissivity is calculated along the downstream regions of the identified shock structures (i.e., from the shock center to the post-shock cell).
  The projected radio flux map obtained in this way is smoothed to match the beam size of the VLA-C observation \citep[\(15''\times15''\), ][]{Botteon2016}.
  We refer readers to \cite{Hong2015} for more details on the individual steps in estimating the radio emission. 
  
\section{Simulation Result}
\label{sec:res}

\begin{figure*}
    \centering
	\includegraphics[width=2\columnwidth]{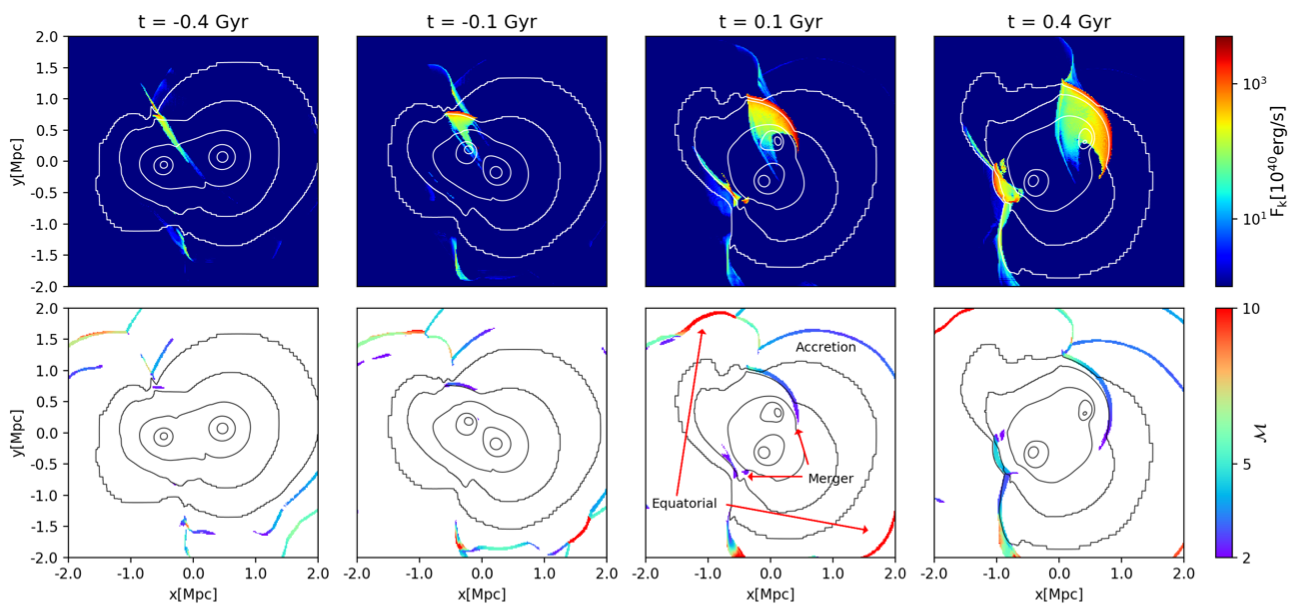}
    \caption{Integrated kinetic energy flux $F_{\rm k}$ (top) and Mach number $\mathcal{M}$ (bottom) distributions. Overlayed
    are the X-ray surface brightness contours (Figure~\ref{fig2}).
    See text for the shock identification details. $F_{\rm k}$ is obtained by multiplying
    the kinetic energy flux $f_{\rm k}$ to the shock surface area and integrating the result along the LOS direction. The Mach numbers $\mathcal{M}$ in the bottom panel are derived by taking the average along the LOS direction within the central $\Delta z=600$~kpc slice.
    The $F_{\rm k}$ and Mach number distributions trace the energy budget and efficiency of radio emissions, respectively. Note the concentration of $F_{\rm k}$ mostly along the merger shock edges.}
    \label{fig3}
\end{figure*}

We present the evolution of the observable features in our simulation at different merger stages.
First, we describe the results from the reference run with the comparison between X-ray morphology and shock structures (\textsection\ref{sec:evolution}).
Then, we discuss detectability of radio emissions in our simulation (\textsection\ref{sec:radiorelic}).
Hereafter, we use the time $t$ elapsed since the first closest passage to denote the merger epoch.

\subsection{Evolution of Observables in the Off-axis Collision Simulation}
\label{sec:evolution}
\subsubsection{X-ray observables}
Figure \ref{fig2} shows the time evolution of our simulated off-axis cluster merger presenting both projected X-ray surface brightness and temperature distributions.
Overall, the result is consistent with a general expectation from previous studies \citep[e.g.][]{Ricker2001}. 
The X-ray luminosity of the system rises as they approach, peaks at the closest passage ($t=0\,\rm Gyr$), and decreases afterwards.
The temperature in the intermediate region between the two clusters increases, as the two clusters approach each other. The maximum temperature reaches up to $>10\,\rm keV$ when the two clusters are near the impact ($t=-0.1\,\rm Gyr$). 
Because of the large pericenter distance ($\mytilde500$~kpc), the two clusters maintain the cool cores after the impact.
We note that temperature folds exist near the density cutoff regions although we employ a large cutoff radius (yellow arrow in Figure \ref{fig2}).
Thus, interpretation of the temperature features beyond the cutoff radius should use caution.
%Apart from these temperature folds, our projected temperature map at $t=0.1$ or 0.4~Gyr resembles the observed temperature map (middle panel of Figure \ref{fig1}).
Apart from these temperature folds, both the X-ray luminosity ($L_{0.1-2.4\rm\,keV}\sim1.3\times10^{45} \rm erg~s^{-1}$) and the projected temperature of the intermediate region after the closest passage ($t=[0.1,0.4]\rm\,Gyr$) agree with the observed level  $L_{0.1-2.4\rm\,keV}=1.5\times10^{45} \rm erg~s^{-1}$ \citep{Botteon2016} and $T=11.03\pm1.7\rm\,keV$ [region D in Table 2 of \cite{Hallman2018}].
This agreement in luminosity must be interpreted with caution, however, because it depends on how exactly one models AGN feedback and radiative cooling, both of which are not included in the current study.

We note that the ``S"$-$shape feature of the central high-temperature region seen at or near the closest encounter ($t=\pm0.1$~Gyr) is not clear in the observed temperature map (Figure~\ref{fig1}).
Assuming that the S/N value of the observed temperature map is sufficient, we suspect that the difference may be attributed to the
intrinsic inhomogeneity of A115, viewing angle, and/or the lack of numerical instability, all of which can blur the boundary of the ``S''$-$shape feature.

For both clusters, cometary tails can be identified from tapering surface brightness ($\Sigma_{\rm Xray} \gtrsim 10^{-4}\rm\,erg~s^{-1}~cm^{-2}$) as well as low temperature regions ($T\lesssim 2\rm\, keV$, Figure \ref{fig2}).
These features start to develop when the two clusters are near the core passage ($t=-0.1\,\rm Gyr$) and become more distinct at $t=0.1\rm\,Gyr$.
The direction of the tails traced by the surface brightness are anti-parallel to the initial velocities.
However, at $t\sim0.4\,\rm Gyr$ after the core passage,
the tails are somewhat shortened with their directions nearly perpendicular to the initial velocities. 
This X-ray tail rotation is known to take place as the gas that was ram-pressure stripped earlier begins to fall back to the cluster potential with a non-zero angular momentum.
This results in a ``slingshot'' feature in the X-ray images \citep{sheardown2019} that shows a morphology different from the ram-pressure stripping tail
observed at $t=0.1$~Gyr.
Remember that at this epoch $t\sim0.4\,\rm Gyr$ the two cluster are still moving away from each other (They reach their apocenters much later at $t\sim2.0$~Gyr) and the orientation of the $t\sim0.4$~Gyr velocity vectors are within ~20$^{\circ}$ from that of the collision axis.
The time evolution of the sub-cluster X-ray tail orientation 
in comparison with the velocity vector is further discussed with shock structures in \textsection\ref{sec:xraytail}.

\begin{figure}
    \centering
	\includegraphics[width=\columnwidth]{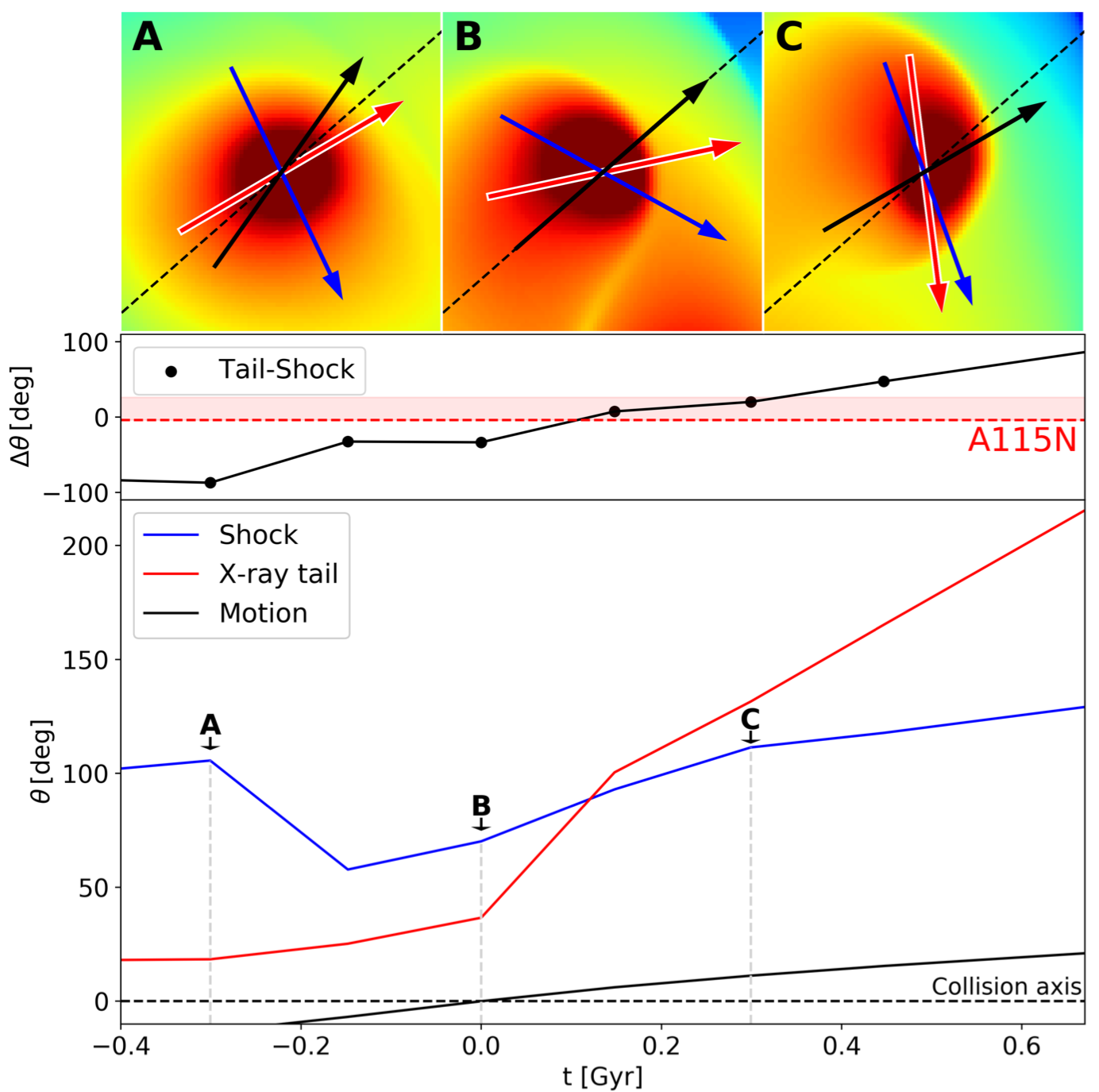}
    \caption{Orientation of the shock, X-ray tail, and dark matter halo velocity. We display the results for the sub-cluster (A115N analog).
     The top panel shows the X-ray surface brightness map at the selected epochs with the various arrows depicting the orientations of the shock (blue), X-ray tail (red), and halo velocity (black); we use the same color scheme in the bottom panel. 
    The collision axis (dashed) is defined to be the velocity vector at the closest passage.
    The middle panel shows the angle between the orientations of X-ray tail and shock. The red dashed line illustrates the observed value at $-4 ^{\circ}$ whereas the shaded region represents
    the systematic uncertainty ($\mytilde30 ^{\circ}$).
    The bottom panel displays the time evolution of the orientation of the observed features with respect to the collision axis.
    }
    \label{fig4}
\end{figure}

Dissociation between dark matter halo and gas is often reported in cluster mergers \citep[e.g.,][]{Clowe2006,Dawson2012}. The gas-mass offset is consistent with our expectation in our cold dark matter (CDM) paradigm because the plasma is subject to coulomb forces whereas the dark matter is believed to be collisionless. In off-axis collisions, the offset is expected to be smaller and decrease with impact parameter.
Both mass peaks of A115 from our WL analysis are  well-aligned with the X-ray peaks as shown in Figure \ref{fig1}. From the current hydrodynamical simulation,
we confirm that no significant offsets between the dark matter and x-ray peaks are present during the period $t=[-0.1,0.4]$~Gyr.

\subsubsection{Merger shocks and X-ray tail alignments}
\label{sec:xraytail}
The shock structures found in the simulation results are displayed in Figure \ref{fig3} using the integrated kinetic energy flux ($F_{\rm k}$, top panel) and the average Mach number in the central $\Delta z=600\, \rm kpc$ slice (bottom panel).
The integrated kinetic energy flux, $F_{\rm k}$ is obtained by integrating $f_{\rm k}\times A_{\rm sh}$ along the line-of-sight direction, where $A_{\rm sh}$ is the shock surface area.

As shown by \cite{Ha2018}, the shock layers (see bottom panel of Figure~\ref{fig3}) can be categorized into three groups: 1) the merger shocks propagating along the merger axis in front of the X-ray cores, 2) the equatorial shock with high Mach numbers ($\mathcal{M} >10$) expanding in a direction perpendicular to the merger axis, and 3) the spherical accretion shock at the outskirt of the cluster\footnote{Our accretion shocks possess low Mach numbers compare to those in cosmological simulations \citep[e.g., ][]{Ryu2003} because of the hot background gas and insufficient time for infall.}.

Among the three shock groups, the integrated kinetic energy flux, $F_{\rm k}$ is dominated by the merger shocks and reaches the highest values along the front edge (see top panel of Figure~\ref{fig3}). The kinetic energy flux of the northern merger shock extends more than $1\rm\,Mpc$ and diminishes precipitously at the eastern density cutoff.
As the kinetic energy flux $f_{\rm k}$ powers the radio emission, we can expect that the radio emission will also be dominated by the merger shocks and become brightest along the edge of the sub-cluster merger shock. A scrutiny of the eastern side of the northern merger shock reveals a sudden increase in the Mach number toward the eastern edge. The feature is attributed to our assignment of a low temperature to the cells in the density cutoff region.
Considering the small temperature gradient of the $\beta$-profile, we think that this artificial temperature drop at the density cutoff may lead to the overestimation of the Mach number (from $\mytilde4$ to $\mytilde7$), which can in turn non-negligibly boost the acceleration efficiency \citep{Kang2013}.
Also, this artifact may result in the premature truncation of the predicted radio relic and overestimation of the radio flux on the eastern side 
(see \textsection\ref{sec:radiorelic}).

Understanding how the orientations of the X-ray tails and merger shocks evolve at different stages of an off-axis merger with respect to the collision axis and the halo velocity vector is one of the key goals of the current study. From visual inspection, one can notice that, unlike the orientation of the X-ray tail of the sub-cluster, the orientation of the sub-cluster merger shock does not vary greatly during the $t=[-0.1,0.4]$~Gyr period. %compared to 
In order to discuss the orientation evolution quantitatively, we use the second moments of the features to evaluate the orientation angle. Employing a proper window function is needed in this calculation because the orientation depends on the scale (in particular for the X-ray tail). For the integrated kinetic energy flux, we consider the data where the values are within the top $0.1\%$, which defines an ellipse whose major axis is $\mytilde1$~Mpc.
The threshold is chosen to be
$>2\times10^{-4}\, \rm erg~s^{-1}~cm^{-2}$ for the X-ray surface brightness. The resulting major axis of the ellipse is $\mytilde300$~kpc.
We determine the cluster motion from the average velocity vector of the dark halo.
In Figure~\ref{fig4} we display the orientation angle as a function of time since the impact.

As qualitatively described before, the direction of the X-ray tail (red arrow in the top panel and red solid line in the bottom panel of Figure~\ref{fig4}) is nearly aligned ($\mytilde30^{\circ}$) with the collision axis at the time of the impact (epoch  {\bf B}). However, its deviation from the collision axis quickly increases in a short period of time and reaches $> 90^{\circ}$ at epoch ${\bf C}~(t=0.3~\mbox{Gyr})$. Because of the large collision velocity, the sub-cluster velocity vector (see the black solid line in the bottom panel of Figure~\ref{fig4}) remains well-aligned with the collision axis until $t=0.6~\mbox{Gyr}$ within $\mytilde20^{\circ}$.

As expected, the orientation of the merger shock is approximately perpendicular to the collision axis from the onset of the merger (epoch ${\bf B}$) and changes only by a small amount throughout the merger duration $t=[-0.2,0.6]$~Gyr investigated here (see the blue solid line in the bottom panel of Figure~\ref{fig4}) \footnote{The shock orientation at epoch ${\bf A}$ is determined by the morphology of the equatorial shock rather than the merger shock, which  is not fully developed yet. See the first column of Figure \ref{fig3}.}. Between epochs {\bf B} and {\bf C}, the shock orientation also undergoes a rotation, which however is much smaller than that of the X-ray tail.

Again, this quantitative analysis confirms that the orientation of the X-ray tail is not a reliable indicator of the sub-cluster motion in an off-axis merger, as already demonstrated by \cite{sheardown2019}.
On the other hand, merger shocks, being relatively insensitive to the stage of the merger, provide more reliable information on the direction of the collision axis.
%On the other hand, merger shocks, relatively insensitive to the stage of the merger, provide more solid information on the direction of the collision axis.

The difference in the rotation rate between the X-ray tail and the merger shock gives an epoch where the orientations of the two structures align ($t\sim0.1$~Gyr, see the middle panel of Figure~\ref{fig4}). Since we witness a similar tail-shock alignment in the current A115 observation, the issue deserves a further test.
One can imagine a number of parameters affecting the exact alignment epoch. However, here we mainly focus on its dependence on the choice in our window function. 
Lowering the threshold of the X-ray tail selection gives a larger area for the evaluation of the orientation. As one can expect from the X-ray map, it decreases the angle between the tail orientation and the collision axis.
Specifically, when we lower the threshold by a factor of two from $2\times10^{-4}\, \rm erg~s^{-1}~cm^{-2}$ to $1\times10^{-4}\, \rm erg~s^{-1}~cm^{-2}$, the decrease is $\mytilde30^{\circ}$ when averaged over different epochs. However, the rotation ``rate" remains similar and thus the total amount of the orientation change $\Delta \theta$ is also similarly large ($\sim146\pm87^{\circ}$) even with this a new window function choice\footnote{With this lower limit, the compressed gas in the central region is occasionally selected as a X-ray tail and additional measures should be taken to correctly measure the shape of the X-ray tail.}. We adopt this window function-dependent change $\mytilde 30^{\circ}$ as a systematic uncertainty in our determination of the orientation angle.
For comparison with the observation, we applied the same procedures to our X-ray and radio data except that we masked out the point sources. Our analysis shows that the observed alignment occurs $0.1-0.4$~Gyr after the collision.

\begin{figure*}
    \centering
	\includegraphics[width=2\columnwidth]{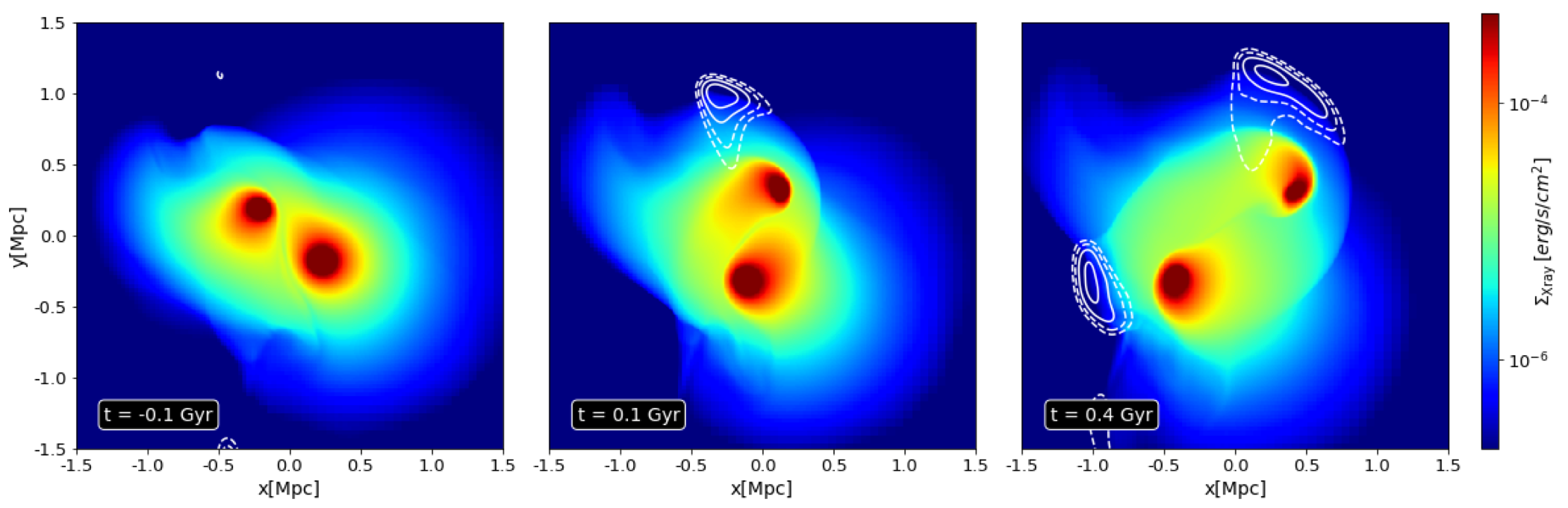}
    \caption{Projected radio flux map. Colors are the X-ray surface brightness maps in Figure \ref{fig2}. Contours represent the projected radio flux, which
    is smoothed with the beam size ($15''\times15''$) of VLA C configuration \protect\citep{Botteon2016} and amplified by a factor of 10 to approximately match the observed level. 
    The solid (dashed) lines show the 3 and $6\sigma$ (1 and $2\sigma$) levels, where
    1$\sigma$ is $70\,\rm \mu Jy~beam^{-1}$.
    }
    \label{fig5}
\end{figure*}
\begin{figure}
    \centering
	\includegraphics[width=\columnwidth]{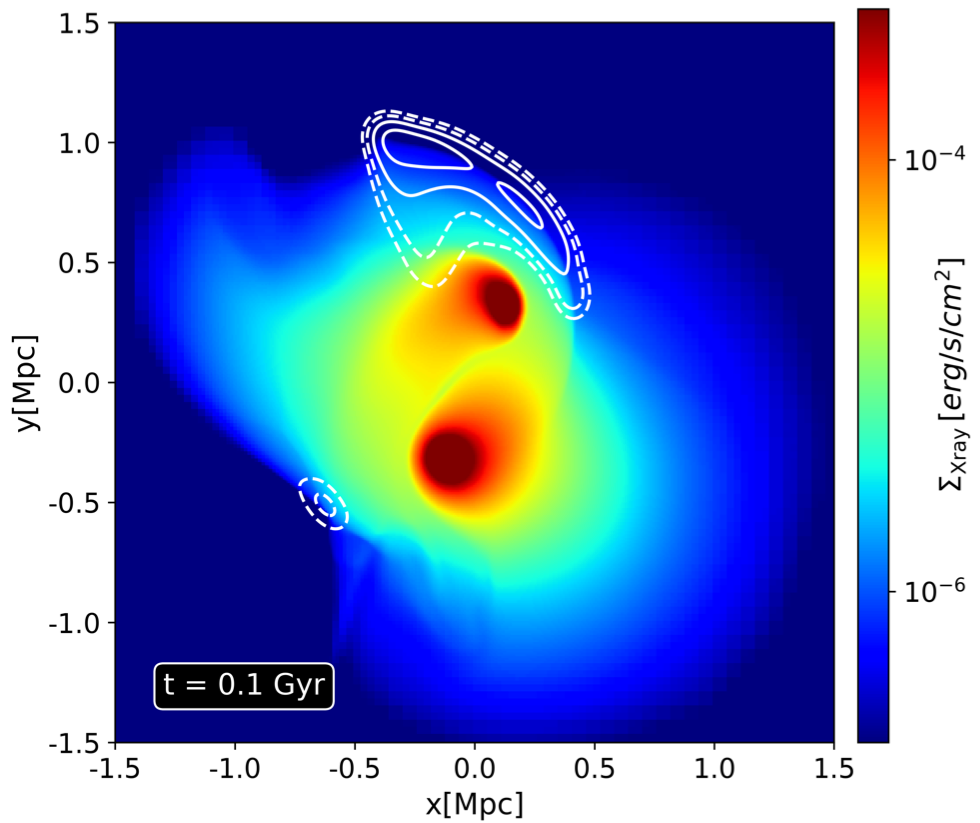}
    \caption{Same as Figure~\ref{fig5} except that here we remove the arbitrary $10\times$ upscaling and instead boost the radio flux with the stronger magnetic field ($3\,\rm\mu G$) and enhanced dissipation efficiency. The enhanced dissipation efficiency is modeled based on the result of \protect\cite{Kang2013}, where the pre-existing CR electrons are assumed to follow the relation $P_{\rm CR}=0.05P_{\rm th}$.}
    \label{fig6}
\end{figure}

\subsection{Radio relics in the off-axis collision}
\label{sec:radiorelic}

We find that the projected radio flux based on the procedure summarized in \textsection\ref{sec:mockradio} is lower than
the noise level $1\sigma=70\,\mu\rm Jy~beam^{-1}$ of the VLA-C observation \citep{Botteon2016}.
Therefore, in our presentation (Figure~\ref{fig5}) we arbitrarily scale up the radio flux by a factor of 10 (see \textsection\ref{sec:reacceleration} for discussion). Readers are reminded that the radio flux contours in Figure~\ref{fig5} and hereafter are displayed with solid (dashed) lines for the  3 and 6 $\sigma$ (1 and 2$\sigma$) levels.

Figure \ref{fig5} shows the northern and southern radio relics along the edge of the simulated merger shocks seen in Figure \ref{fig3}.
The orientation of the northern radio relic roughly agrees
with that of the sub-cluster merger shock, however, with some notable differences.
At the $0.1<t<1.0\rm\, Gyr$ epoch, during which the peak of the radio flux is $>3\sigma$, the orientation of the relic remains nearly constant ($\Delta \theta \lesssim 3 ^{\circ}$)\footnote{We determine the orientation by fitting an ellipse to the solid contours.}. The merger shock undergoes a larger change ($\Delta \theta \lesssim 20 ^{\circ}$) in orientation.
This is because the part of the sub-cluster merger shock ahead of the sub-cluster has a low Mach number ($\mytilde2$, see Figure \ref{fig3}), which results in a low radio emissivity.
The nearly constant radio relic orientation angle  $\mytilde62 ^{\circ}$ with respect to the collision axis is similar to the merger shock orientation at $t\sim0~\rm\,Gyr$. Thus, our simulation result illustrates that the orientation of the observed radio relic might serve as a reliable indicator of the merger axis, insensitive to the stage of the merger.

Our simulated relic is somewhat shorter than the observed radio relic. One may suspect that this may arise from the eastern side truncation mentioned in \textsection\ref{sec:shock}. However, one should also remember that the Mach number in this region is likely to be overestimated because of the same truncation effect (thus the relic on this side is estimated to be brighter). Correcting for this Mach number overestimation would result in a factor of $\mytilde4$ reduction in radio flux, providing a smoothly decreasing radio flux toward the eastern boundary before the truncation. 

\subsubsection{Enhancement of radio relics
through stronger magnetic fields and fossil electrons}
\label{sec:reacceleration}

As mentioned before, the reconciliation
in radio flux between our simulation and the observation requires a factor of 10 boost in the simulated radio flux. This supports the well-known claim that the acceleration of electrons from the thermal
pool (via DSA) is too inefficient to 
explain the observed luminosity of typical radio relics \citep[e.g.][]{Botteon2019}.
One suggestion to resolve the issue is the so-called ``re-acceleration model", which hypothesizes that fossil relativistic electrons from radio galaxies are re-accelerated by the DSA mechanism. The theory has been gaining support with the discovery of a growing number of clusters hinting at such radio galaxy-relic connections \citep[e.g.,][]{vanweeren2017}.

Perhaps, A115 can also serve as such an example because there are a few confirmed radio sources 
blended with the radio relic \citep{Harwood2015}. In this paper, we test the enhancement of the radio relic flux by assuming
a factor of 3 stronger magnetic field ($B=3\,\mu\rm G$) with the availability of significant fossil electrons. We implement the impact of fossil electrons through a higher dissipation efficiency $\eta_{\rm eff}(\mathcal{M})$. As shown in equation \ref{eq:fk}, the stronger magnetic field increases the overall radio emissivity, whereas the modified dissipation efficiency assuming the pre-existing CR (i.e. $P_{\rm CR} = 0.05 P_{\rm th}$) enables the weak shocks ($\mathcal{M}\sim2$) to accelerate particles \citep{Kang2013}.

We display the resulting radio flux map at $t=0.1\rm\,Gyr$ in Figure~\ref{fig6}. 
With the aforementioned recipe, this new
radio flux matches the observed level without the
arbitrary $10\times$ upscaling employed in Figure~\ref{fig5}.
Moreover, the northern relic is stretched further
to the west than the previous case because the enhanced dissipation enables weak shocks in the western region to emit radio. This increase in length by a factor of two also better matches the observation.
In the eastern shock region, the simulated radio signal remains high ($>6\sigma$) even after 
we correct for the overestimated Mach number, which
implies that potentially the northern radio relic could extend further beyond the truncation and 
reach the observed length of $\sim2\rm\,Mpc$ if there were no boundary effect.

Figure \ref{fig6} shows that the southern main-cluster merger shock too becomes detectable.
The absence of the corresponding signal in the current radio observation may be attributed to the lack of radio point sources in the southern region if indeed fossil electrons are mainly responsible for the enhancement of the radio relic flux.

We adopt a uniform magnetic field strength ($1~\mu G$) to estimate our radio emissivity.
Considering the rather turbulent nature of the ICM magnetic fields \citep[e.g.,][]{Roh2019}, we think that it is probably unrealistic to assume 
such homogeneity over the $\mytilde2~\rm Mpc$ scale.
Moreover, if this radio relic was formed by re-acceleration of low energy cosmic-ray electrons in a dead radio jet,
which has been mixed with entrained ICM, the fossil plasma may possess remnant magnetic fields as well.
Detailed modeling of these features is beyond the scope of the current paper.

\begin{figure*}
    \centering
	\includegraphics[width=2\columnwidth]{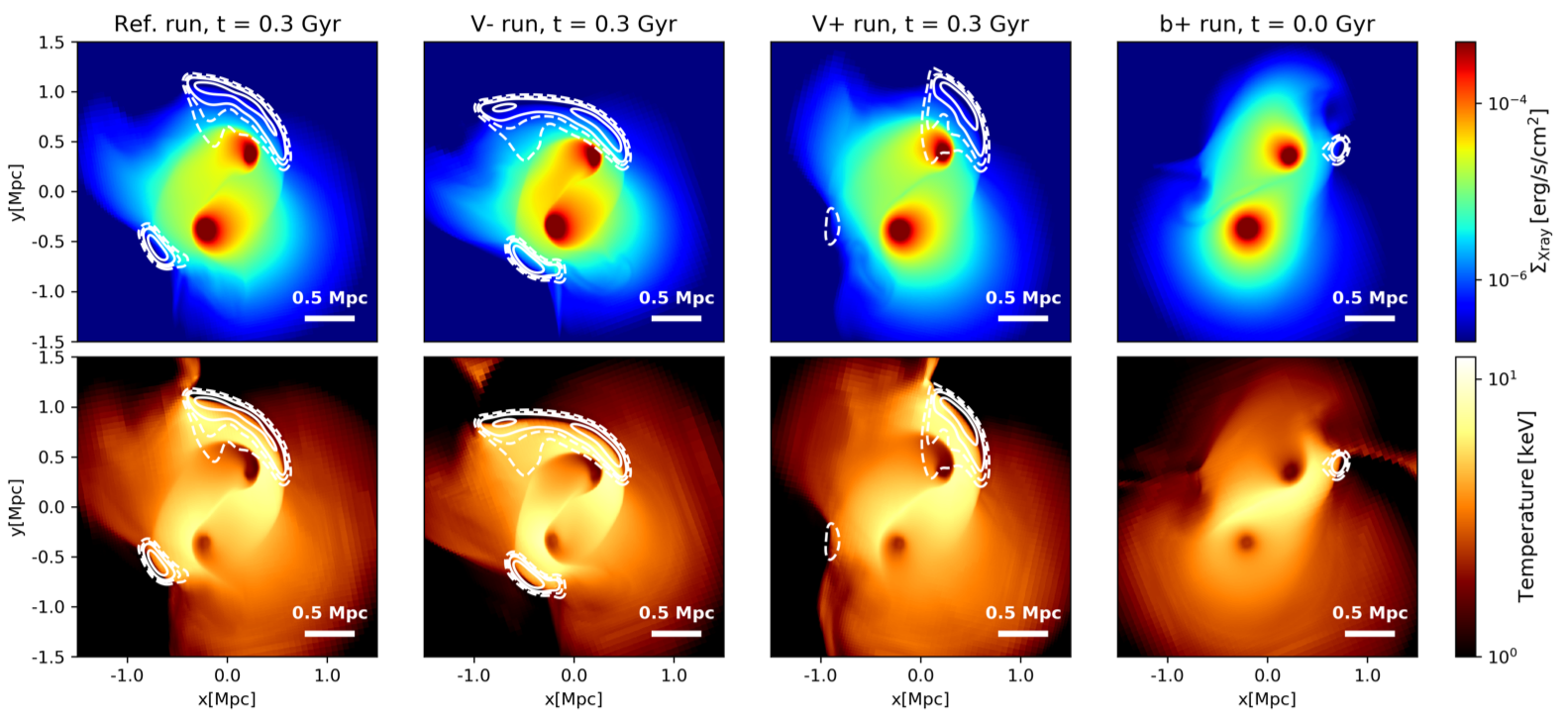}
    \caption{Projected radio flux map overlaid on the X-ray surface brightness (top) and temperature map (bottom) for the runs with different collision setups. The collision setup is labeled on top of each column. The snapshot epoch shown here is selected when the distance between the clusters becomes comparable to the observation $\mytilde900$~kpc after the first passage. The radio flux is boosted with the same magnetic field strength and efficiency used in Figure \ref{fig6}.
    }
    \label{fig7}
\end{figure*}

\section{Discussion}
\label{sec:dis}
\subsection{Effect of Collision Parameters}
\label{sec:effectofcollision}
%As mentioned earlier, 
In addition to the reference run, we test three additional cases, where the collision velocity
is lower ($V-$ run), the collision velocity is 
higher ($V+$ run), and the pericenter distance is larger ($b+$ run).
Figure \ref{fig7} presents the resulting X-ray density and temperature maps overlaid with radio contours enhanced through the same recipe explained in \textsection\ref{sec:reacceleration}.
Because these three simulations are now at different stages for the same simulation time $t$, we compare the epochs, when the distance between the clusters is $\mytilde900 \,\mathrm{kpc}$ after the core passage.

On large scales, the three simulations in Figure \ref{fig7} share similar distributions in density and temperature to those of the reference run, featuring cool cores and hot intermediate regions.
On small scales, a few notable differences are present.
First, the X-ray tails become more prominent when the collision velocity is lower.
We attribute this to the smaller cluster separation at the impact, which results in the stronger drag force.
Second, the length of the radio relic also increases for the lower velocity because more space is accessible to the merger shock.

The $b+$ run is selected at  $t=0~\rm\,Gyr$ since the smallest separation for this case is already comparable to the observed value $\mytilde900~\rm\,kpc$.
The simulated X-ray map shows a distinct ram-pressure stripping tail, which approximately corresponds to the merger scenario proposed by \cite{Hallman2018}.
The triangular shape of the cometary tail and
the hot temperature in the intermediate region resembles the observed features.
Nevertheless, the anticipated merger shock,
even after we consider re-acceleration of cosmic rays,
is too weak to produce the observed radio flux.
Also, the expected peak location of the radio emission is different from the observed position.

\begin{figure}
    \centering
	\includegraphics[width=\columnwidth]{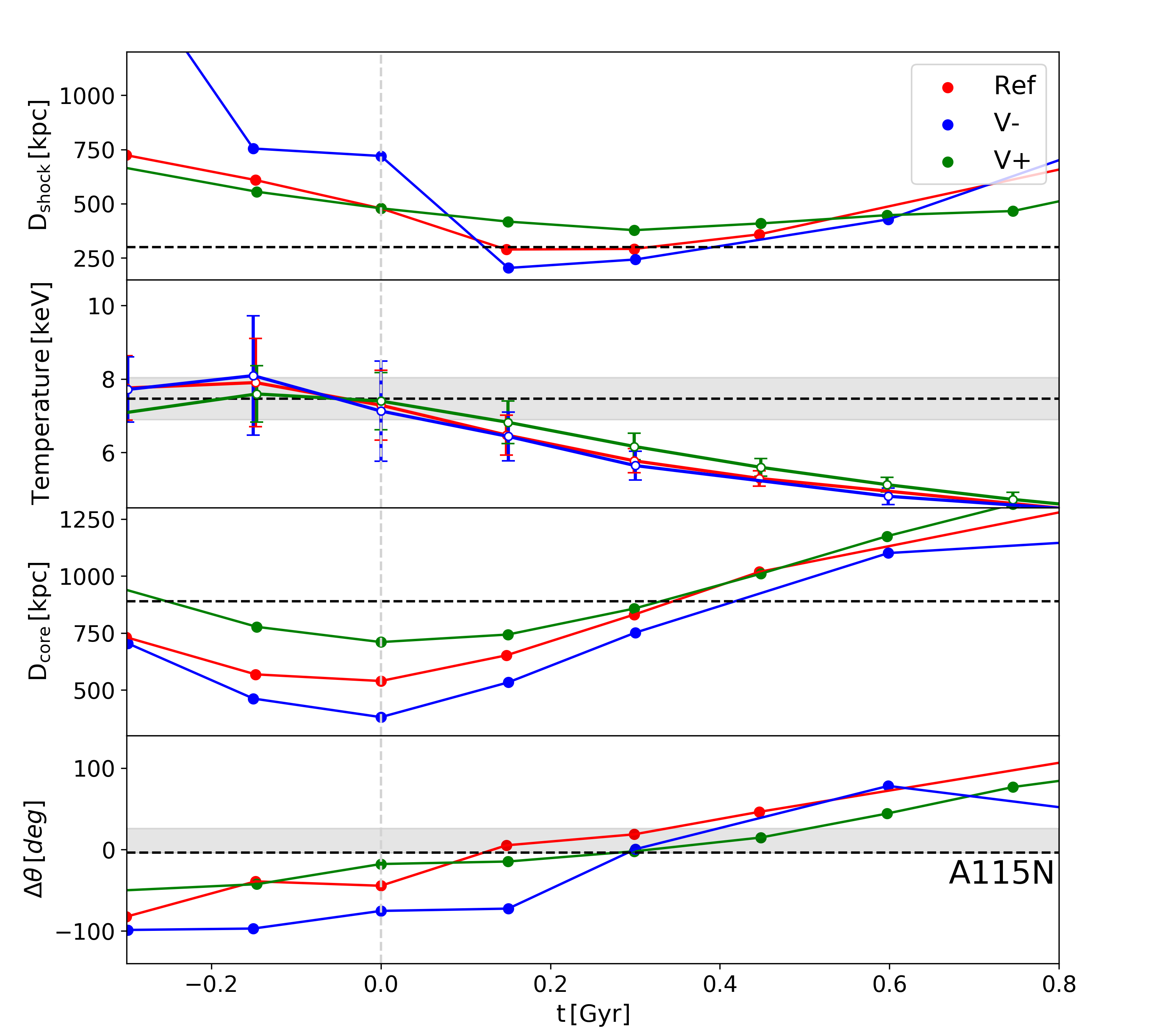}
    \caption{Time evolution of the observable features.
    The dashed line is the observed value whereas the solid lines are for the different simulation runs.
    From top to bottom, the distance between the sub-cluster merger shock and X-ray core $D_{\rm shock}$, the intermediate region temperature, the distance between the two cluster cores $D_{\rm core}$, and the angle $\Delta\theta$ between the sub-cluster X-ray tail and  merger shock are presented. The shaded regions in the temperature ($\Delta\theta$) panel represents the statistical (systematic) uncertainty. See text for the description of the systematic uncertainty in the $\Delta\theta$ measurement due to the window function choice.}
    \label{fig8}
\end{figure}

The radio relics in Figure~\ref{fig7}, except for the $b+$ run, all show an apparent alignment with the X-ray tails.
For the three simulation runs with a distinct radio relic, we compare the following four physical properties with the observations: the distance between the sub-cluster merger shock and X-ray core $D_{\rm shock}$, the temperature in-between the clusters, the distance between the two X-ray cores $D_{\rm core}$, and the alignment between the sub-cluster X-ray tail and the sub-cluster merger shock $\Delta\theta$ in Figure \ref{fig8}.

The distance to the merger shock $D_{\rm shock}$ is computed using the separation between the kinetic energy flux ($f_{\rm k}$)-weighted center of the merger shock and the X-ray core for the simulations. For the observation, we use the radio flux-weighted center of the radio relic and the X-ray center of A115N. The radio point sources are masked out when computing the flux-weighted center of the observed radio relic. 

The angle $\Delta\theta$ is measured in the same way as in the creation of Figure \ref{fig4} for both simulations and observations. The systematic uncertainty due to the window function choice is  considered to be $+30^{\circ}$ as before.

The central temperature in the simulation is computed using equation \ref{eq:Temp} with the cells within the $400\rm\,kpc \times 200\rm\,kpc$ box positioned between the clusters with the shorter side aligned parallel to the axis connecting the two clusters.
Within the box, we averaged the projected temperature to derive the central temperture of the simulation. To determine the temperature
from the observation, we first extract the X-ray spectrum from the equivalent region  ($\mytilde2'\times1'$) using {\tt ciao}-4.11;
we merged the X-ray spectra from the observations  3233, 13458, 13459, 15578, and 15581 and binned the counts in such a way that each energy bin has a signal-to-noise ratio of 5. 
We fit an absorbed {\tt MEKAL} model \citep{Mewe1985,Kaastra1993,Liedahl1995} to the binned spectrum using the $0.5-7 \rm\, keV$ energy range and determine the temperature to $7.46\pm0.57\rm\,keV$. We verify that a highly consistent result is obtained when we use the Astrophysical Plasma Emission Code \citep[APEC;][]{Smith2001} plasma model instead of the MEKAL model.

The shock distance (measured from the subcluster X-ray core) $D_{\rm shock}$ slowly decreases until it reaches the minimum at $t\mytilde0.1-0.3$~Gyr (Figure~\ref{fig8}).
Although the exact values differ among the simulation setups, they all are similar to the observed value (dashed).
After this epoch, $D_{\rm shock}$ gradually increases with a moderate slope, which indicates the launch of the merger shock \citep{Sheardown2018a}.

As one can expect, the temperatures in-between the clusters show a gradual decrease after they peak at the impact. The peak temperatures are similar to the observed value. Nevertheless, one should not use this temperature comparison to constrain the merger scenario/epoch because a few keV difference in temperature can be easily resolved by ``fine-tuning" simulation setups, which is beyond the scope of the current paper.

The distances between the cores $D_{\rm core}$ match the observed value at $t\mytilde0.3$~Gyr for the reference and $V+$ runs and at $t\mytilde0.4$~Gyr for the $V-$ run.

Considering the systematic uncertainty due to the window function, we can estimate that the alignments between the X-ray tail and merger shock happen during the $t=[0,0.4]\rm\,Gyr$ period.

In summary, our simulations show that the three observed properties of A115 can be reproduced at $t\mytilde 0.3$~Gyr after the core impact.

\begin{figure}
  \centering
	\includegraphics[width=\columnwidth]{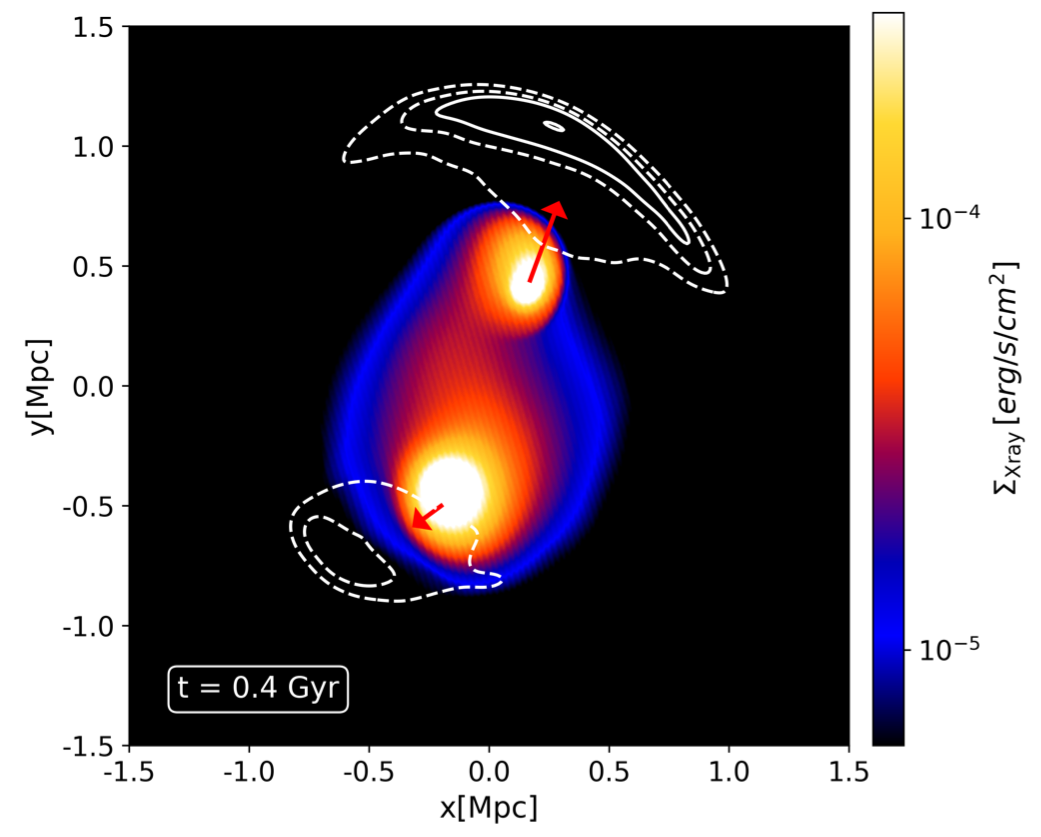}
    \caption{Impact of viewing angle on the X-ray surface brightness and radio emission.
    We display the case where the merger axis is tilted by $\mytilde30^{\circ}$ with respect to the plane of the sky.  Compared to the one used in Figure \ref{fig2}, we apply a stretched color scheme to match the range used in Figure \ref{fig1}. The projected motion of the clusters is marked with red arrows. The velocity of the sub-cluster (main cluster) is $\mytilde800\rm\, km~ s^{-1}$ ($\mytilde300\rm\, km~ s^{-1}$). }
    \label{fig9}
\end{figure}

%\section{Discussion}
%\label{sec:dis}
\subsection{Viewing Angle of A115}
\label{sec:viewing}
Our simulations described in \textsection\ref{sec:evolution} and \textsection\ref{sec:effectofcollision} show that many observed features such as the morphology, location, and orientation of the X-ray tail and radio relic can be reproduced by an off-axis merger $\mytilde0.3$~Gyr after the core passage, which favors a ``slingshot" mechanism rather than a ram-pressure stripping for the origin of the observed X-ray morphology.
These simulations assume a plane-of-the-sky merger,
which is motivated by the small line-of-sight velocity difference between the member galaxies of A115N and A115S \citep{Golovich2018}.
However, detailed analysis of 
the velocity structure near the cluster cores with the cluster galaxies and gas hints at the possibility that the A115 merger 
may have a significant LOS velocity component \citep{Barrena2007,Kim2019}.
For example, \cite{Kim2019} show that the relative velocity between the two BCGs of A115N and A115S is large ($853\pm6\rm\,km~s^{-1}$).
They also demonstrate the robustness of this result by including in their analysis the member galaxies
within $250\rm\, kpc$ radius from each BCG, although the significance is somewhat lower ($838\pm549 \rm\,km~s^{-1}$).
Using X-ray spectra, \cite{Liu2016a} report an extremely large velocity difference
($\mytilde 4600\pm1100\rm\,km~s^{-1}$) between A115N and A115S.
Although we think that more studies are needed to settle the A115 LOS velocity issue, here we discuss how the results change when we assume a non-zero viewing angle.

If we hypothesize that the relative LOS velocity between the BCGs ~($\mytilde850\rm\,km~s^{-1}$)~ is representative of the LOS velocity difference between A115N and A115S, the observed projected distance $\mytilde 900\rm\,kpc$ between the X-ray cores and the relative LOS velocity can be obtained at $t=0.4\rm\, Gyr$ by rotating the merger axis of the reference run by $\mytilde30^{\circ}$. Figure~\ref{fig9} shows the resulting X-ray density map and the radio relics. 

When rotated by $\mytilde30^{\circ}$, the X-ray morphology shown in Figure \ref{fig9} features an enhanced bridge between the two clusters. Also,
the X-ray tails look somewhat more compact than the unrotated version, more similar to the observation (Figure \ref{fig1}).

The peak radio flux is reduced by $\mytilde80\%$, although the overall flux and morphology still resemble the radio data. The properties described in Figure \ref{fig8} do not change very much because of the small viewing angle; the angle $\Delta\theta$ is $\mytilde 50 ^{\circ}$ and the distance to the shock center $D_{\rm shock}$ from the X-ray core is $\mytilde 500 \rm\, kpc$.
In this rotated projection, the orientation of the radio relics is still nearly perpendicular to the cluster motion whereas the projected X-ray tail becomes further rotated because the selected epoch here (to match the projected distance between the cores) corresponds to a later merger phase.
The rotated temperature map (not included in the current paper) shows a somewhat reduced temperature contrast compared to Figure \ref{fig2}. However, because of the small viewing angle considered here, the overall difference is insignificant.

From this simple experiment, we can reason that an introduction of a small viewing angle ($\mytilde30^{\circ}$) still provides a consistent picture of the merger scenario of A115. However, we believe that a viewing angle significantly larger than this  \citep[e.g.,][]{Liu2016a} cannot reconcile
with the observation. 
A larger viewing angle requires a larger 3-dimensional separation between A115N and A115S. Therefore, we do not expect that any noticeable X-ray tails with the current observed orientations would be produced at 
this large separation (i.e., at the very late phase). Of course, the formation of radio relics matching the current radio data would be even more challenging.

\subsection{Merger Scenario and Caveats}
%\subsection{Caveats}
\label{sec:caveats}

\begin{figure}
    \centering
	\includegraphics[width=\columnwidth]{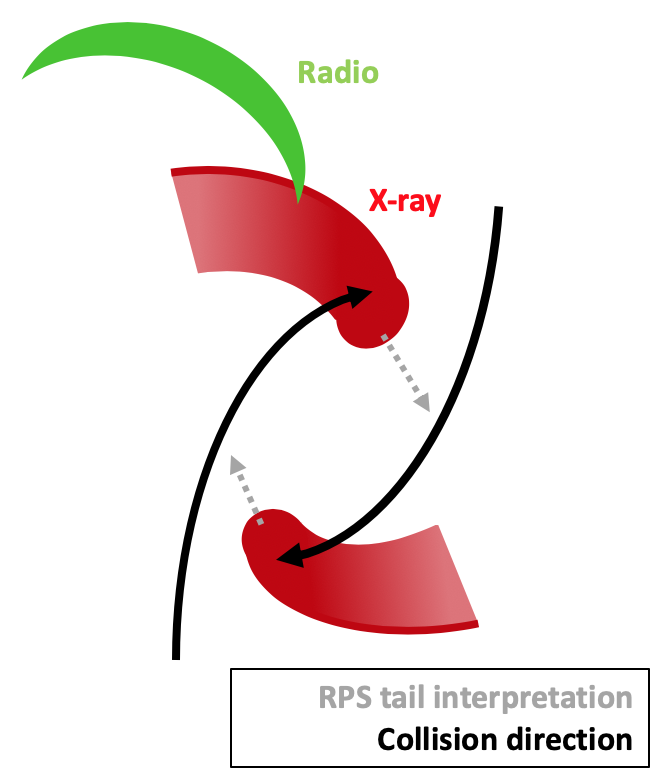}
    \caption{Schematic diagram of the merger scenario of A115 proposed in the current study. The clusters experienced an off-axis collision about $\mytilde0.3$~Gyr ago. They have been following the trajectories depicted with black arrows. The merger shock generated at the impact has travelled along the merger axis and is observed at the current position. The early X-ray tails from the ram-pressure stripping have subsided and now the infalling gas creates the slingshot tails. The dotted gray arrows depict the motion of the clusters in case we interpret the observed X-ray tails as originating from ram-pressure
    stripping.}
    \label{fig10}
\end{figure}

Our simulations show that the observed morphology, orientation, and location of both the X-ray emission and radio relic of A115 can occur during the $t=0.1-0.3$ Gyr period after an off-axis collision before the two subclusters reach their apocenters. In particular, this scenario can explain the parallel alignment of the radio relic and the northern X-ray tail within the conventional paradigm, in which  the relic traces the location of the merger shock.
Although the details change when different initial velocities and impact parameters are selected, this alignment is always reproduced in our experiments with a zero or modest viewing angle.
We present the schematic diagram showing our merging scenario in Figure~\ref{fig10}.

\cite{Golovich2018} suggested that the two subclusters in A115 might be in their returning phase. This scenario is motivated by the direction of the observed X-ray tails which are interpreted as ram pressure features. The main drawback of this scenario is its inability to explain the proximity of the relic to the northern X-ray core. As our simulations show, the on-set of the second in-fall happens at $t=1.8$ Gyr, by which time the merger shock is located well outside the simulation box.

The \cite{Hallman2018} scenario is similar to that of \cite{Golovich2018} in that they attribute the X-ray tails to ram pressure stripping. The difference is that they explain that "the radio relic results from the Fermi acceleration of a relic population of cosmic rays ejected from the
local radio galaxies and advected behind the motion of A115N." Although the authors are implicit regarding the merger phase, the proximity
of the relic requires the relic to be young. Hence, although we agree that the local galaxies might be associated with the radio relic, it is difficult to produce such a fully developed merger shock at this early phase.

%Our simulations have found that the observed alignment between the X-ray tail and the radio relic appears during the post-merger of an off-axis collision, consistently with different setups and a small viewing angle. Our revised merger scenario on A115 can thus be summarized in a schematic diagram in Figure \ref{fig10}. Compare to the past suggestions, our revised scenario explains A115 at an earlier phase than the 2nd infall scenario of \cite{Golovich2018}, which now can easily explain the close separation between the Northern X-ray core and the radio relic. Compare to the ram-pressure stripping tail scenario of \cite{Hallman2018}, the later merger phase of our scenario allows the merger shock to fully develop and generate the radio relic under some plausible assumptions.}

We cannot completely rule out the possibility of the ram pressure stripping tail scenario because this scenario can better reproduce the triangular morphology of A115N X-ray tail (see the b+ run in Figure \ref{fig7}). However, the difficulty in this scenario is that our estimation of the radio emission via merger shock is incompatible with the observed radio emission, in both flux level and the location. This challenge (if one favors a ram-pressure scenario) can be overcome if what we have called the radio relic so far might in fact have a different origin.
For example, as mentioned in \cite{Botteon2016}, energized trails of the old AGN tails [i.e. radio phoenix, \cite{DeGasperin2017}] may mimic a relic-like morphology as observed. Future deeper radio
observations can test this possibility.

Despite the ease with which we can explain the key observed features, our favored merger scenario has some limitations regarding detailed structures.
Our cluster merger is modeled as a collision between the two spherically symmetric halos, which
inevitably produces a smooth merger shock.
On the other hand, the merger shocks in both cosmological simulations and observations show inhomogenous Mach numbers \citep[e.g.,][]{Hong2015} and filamentary substructures \citep[e.g.,][]{Vazza2016, Rajpurohit2018}.
Although the main X-ray and shock properties are determined by the cluster collision itself by and large \citep{Kang2007}, 
real-world inhomogeneities in halo and background properties can modify the simulated features.
For instance, turbulence in the X-ray tail can be generated \citep[e.g.,][]{Donnert2017}.
In our simulation, the turbulent instability along the shear of the X-ray tail
(i.e. Kelvin-Helmholtz instability) is only weakly developed.
As other studies with comparable or larger resolution also show similar features to our results \citep[e.g.,][]{Brzycki2019}, we speculate that perhaps the limited resolution might have
prevented a full development of the instability.
Nevertheless, since our analysis focuses on the X-ray tail near the center, where the instability plays a negligible role \citep[see Figure 2 in ][]{sheardown2019}, we expect our conclusions to remain valid even with the presence of instabilities. 

The X-ray tail in our favored merger scenario is somewhat curved whereas
this feature is not clear in the {\it Chandra} observation.
As mentioned earlier, this morphological discrepancy can be mitigated with a non-zero viewing angle (Figure \ref{fig9}). Obviously, an introduction of contrived inhomogeneous environment might also produce X-ray features more similar to the observation.

%Our simulation shows that a ram-pressure stripping tail produces the triangular morphology of the A115N X-ray tail (see the $b+$ run in Figure \ref{fig7}).
%However, the radio emission from this setup is estimated to be very weak and only marginally detectable at the
%western edge of A115N. According to the  radio relic generation mechanism adopted in this study, both the flux level and the location are highly incompatible with the observed radio emission.
%This challenge (if one favors a ram-pressure scenario) can be overcome if what we have called the radio relic so far might in fact have a different origin.
%For example, as mentioned in \cite{Botteon2016}, energized trails of the old AGN tails [i.e. radio phoenix, \cite{DeGasperin2017}] may mimic a relic-like morphology as observed. Future deeper radio observations can test this possibility.

The model used in our estimation of the radio flux is rather limited, since we adopt a simple DSA prescription for the proton acceleration at parallel shocks \citep{Kang2013}. The shock criticality and obliquity are not considered, because we do not follow the evolution of magnetic fields in our hydrodynamic simulations. Recently, \cite{Kang2019} suggested through particle-in-cell simulations that electrons may be injected to the DSA process only at supercritical ($\mathcal{M}_{\rm s}>2.3$), quasi-perpendicular shocks in high plasma beta ($=P_{\rm g}/P_{\rm B}$) ICM plasma. Future studies using MHD simulations will be useful to quantifying the electron acceleration at ICM shocks driven by off-axis cluster mergers.

%The model used in our estimation of the radio flux is incomplete.
%For example, the dissipation efficiency used in our calculation is based on the CR proton acceleration at the shock  propagating parallel to the magnetic field \citep[i.e. quasi-parallel shock,][]{Kang2013}.
%However, recent studies suggest that obliquity of the shock with respect to the magnetic field can significantly alter the acceleration efficiency of electrons \citep[e.g.,][]{Wittor2017,Kang2019}. Future studies using MHD simulations will be useful to enhance our understanding of the acceleration efficiency in off-axis cluster mergers.

\section{Summary}
\label{sec:con}
A115 is an off-axis binary merger with a number of remarkable features in multi-wavelength
observations. Although the cluster has been a target of quite a few studies,
no convincing merger scenario that explains the observed properties of both the X-ray tail and
radio relic with a coherent scheme has been presented. In this paper, we have studied A115 with high-resolution ($\mytilde 8 \rm\,kpc$) hydrodynamical simulations, leveraging our multi-wavelength observational data.
We design the idealized simulation of the cluster collision in such a way that the observed features are reproduced at the cluster redshift.
Our findings are summarized as follows.

\begin{itemize}
    \item Our simulations show that the observed morphology, null mass-gas offset, orientation, and position of the northern X-ray tail and radio relic can be reproduced by a 2:1 cluster-cluster collision occuring $\mytilde0.3$~Gyr ago with a pericenter distance of $\mytilde500$~kpc and an impact velocity of $\mytilde2,000~\mbox{km}~\mbox{s}^{-1}$, as described in the schematic diagram of Figure \ref{fig10}.
    
    \item The orientation of the X-ray tail can become parallel to that of the radio relic 
    as the early ram-pressure tails have weakened and the infalling gas creates the 
    slingshot tails \citep{sheardown2019}.
    
    \item The orientation of the X-ray tail shows a large variation throughout the merger. We agree with \cite{sheardown2019} that blindly attributing the X-ray tail to  ram-pressure stripping can lead to an incorrect merger phase.
    
    \item Unlike the orientation of the X-ray tail, the merger shock, hence the radio relic, maintains a stable orientation throughout the merger. Therefore, estimation of the collision axis based on the observed orientation of a radio relic is relatively reliable.
    
    \item With DSA alone, the simulated radio flux is lower than the observation by an order of magnitude. This result reconfirms the DSA inefficiency problem. With a stronger magnetic field and enhanced dissipation, our simulation can reproduce the observed strength of the radio remission.
    
\end{itemize}

Our conclusions above can be used to interpret other well-known off-axis cluster mergers such as MACS J1752+4440 \citep[e.g.,][]{VanWeeren2011}, Abell 141 \citep[which is also known as the analog of A115,][]{Caglar2018}, Abell 1644, and some examples mentioned in \cite{sheardown2019}.

Under the re-acceleration hypothesis, the presence of the confirmed radio point sources in the northern side of A115 might have enabled the radio flux in this region to reach the current, observed level. In order to enable further investigation of the radio source-relic connection, more multi-frequency radio observations of A115 are needed. Currently, only a single band ($1.4\rm\,GHz$) radio image has been published. Future low-frequency radio observations and the spectral slope analysis will greatly enhance our understanding of the formation of radio relics and the role of AGNs.

\section*{Acknowledgements}
This study is supported by the programs Yonsei University Future-Leading Research Initiative (2019-22-0056 and 2019-22-0216).
MJJ acknowledges support for the current research from the National Research Foundation of Korea under the programs 2017R1A2B2004644 and 2017R1A4A1015178.
HK was supported by the NRF of Korea through grant 2017R1D1A1A09000567.
DR was supported by the NRF of Korea through grants 2016R1A5A1013277 and 2017R1A2A1A05071429.
TK was supported by the National Research Foundation of Korea under the programs 2017R1A5A1070354 and 2018R1C1B5036146.
We thank Eric Hallman for sharing his figure and Bill Forman, Christine Jones, David Wittman, Sungwook Hong, Craig Lage, and Kyle Finner for useful discussions.
The CosmoSim database used in this paper is a service by the Leibniz-Institute for Astrophysics Potsdam (AIP).
The MultiDark database was developed in cooperation with the Spanish MultiDark Consolider Project CSD2009-00064.

\bibliography{reference}

\appendix
\section{Parameter Test with a Large-Volume Cosmological Simulation}
\label{app1}
  \begin{figure}
    \centering
	\includegraphics[width=0.6\columnwidth]{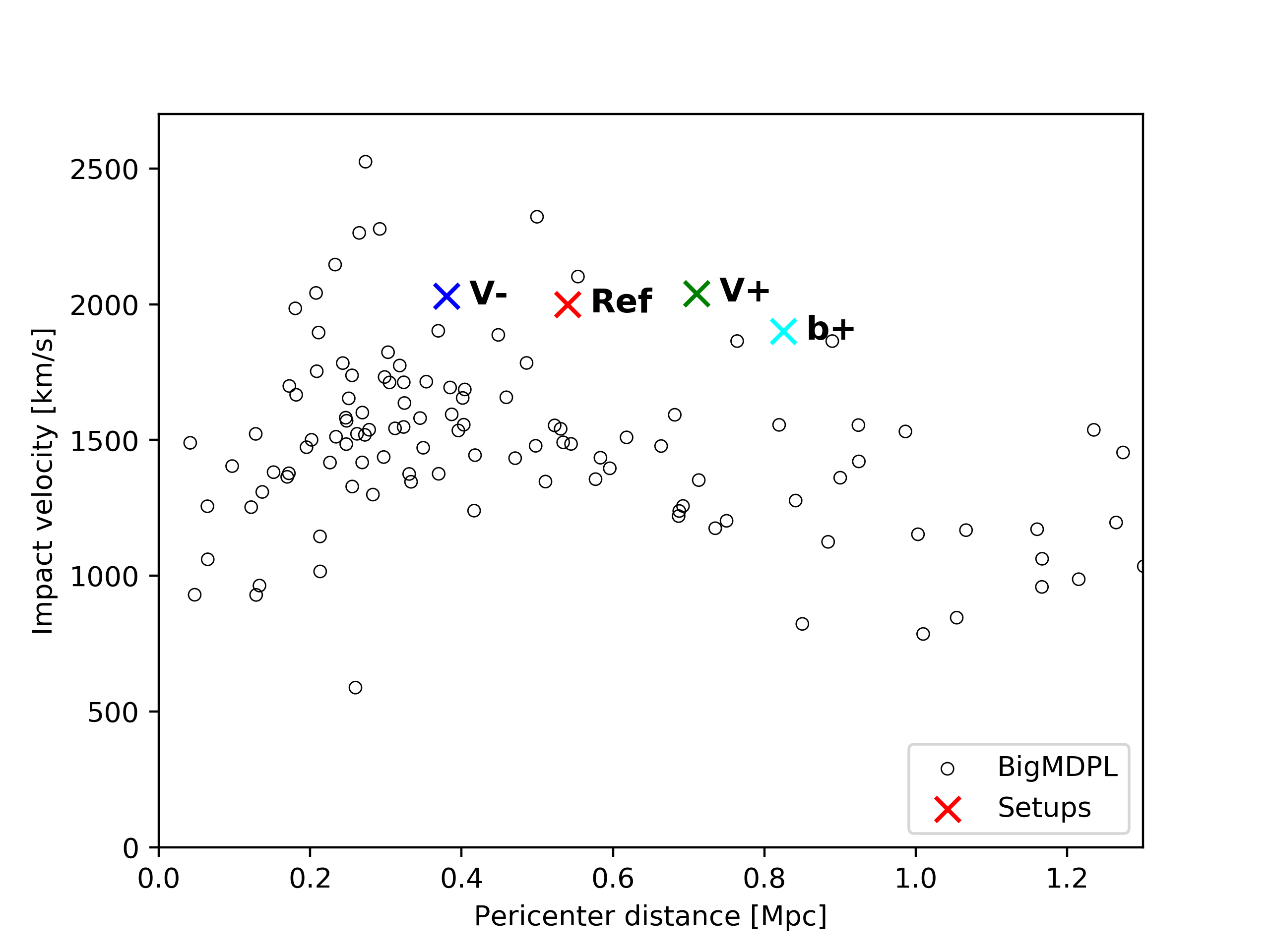}
    \caption{Distribution of the pericenter distance and impact velocity of the cluster mergers at their first passage found in the BigMDPL simulations \citep{Klypin2016}. The cluster mergers are selected if  the total mass is in the $[2\times10^{14}M_{\odot},8\times10^{14}M_{\odot}]$ range and the mass ratio is $<3$.
    The comparison shows that our simulation setups (crosses) are bracketed by the merger properties in the large-volume simulation.}
    \label{fig:app}
  \end{figure}
  Our collision velocity $\mytilde2,000~\mbox{km}~\mbox{s}^{-1}$ slightly exceeds
  the maximum velocity $\mytilde1,800~\mbox{km}~\mbox{s}^{-1}$ derived from the timing argument \citep{Sarazin2002}, which assumes that the merger speed increases through free-fall acceleration until the observed redshift.
  To test the plausibility of our collision scenarios, we compare our simulation setups with the cluster mergers found in the large-volume $(1\, \rm Gpc^3)$ cosmological N-body simulation BigMDPL \citep{Klypin2016} in Figure \ref{fig:app}.
  
  From the samples in this cosmological simulation, we first identify merger candidates
  and track their separations as a function of time. A subsample
  of the merger candidates are selected as mergers when their separations experience  a turnaround. 
   The distance between the clusters at the turnaround and the relative velocity at this epoch is considered the pericenter distance and impact velocity, respectively \citep[see ][for detail]{Wittman2018a}.
  From a total of 73,620 cluster pairs, we select 108 A115-like major mergers 
  if their total mass is bracketed by the interval $[2\times10^{14}M_{\odot},8\times10^{14}M_{\odot}]$ and their mass ratio is $<3$.
  Readers are reminded that the total mass and the mass ratio of A115 from the recent WL analysis are $\mytilde6\times10^{14}M_{\odot}$ and $\mytilde 2$, respectively \citep{Kim2019}. 
  
 Figure \ref{fig:app} shows the distribution of pericenter distance and impact velocity of the cluster mergers found in the BigMDPL simulation. The distribution approximately peaks at
 the impact velocity of $\mytilde1,600\rm\,km~ s^{-1}$, which is consistent with the expectation from the timing argument. 
  Nevertheless, there are cluster mergers with higher impact velocities up to $\mytilde2,500\rm\,km~ s^{-1}$ and the distribution covers the parameter space spanned by our simulation setups.
  The impact velocities of the $V+$ and the $b+$ runs are higher than those of the cluster pair samples with similar pericenter distances. However, the difference is small ($\lesssim 200 \rm\,km~ s^{-1}$). 
  Therefore, the comparison justifies our 
  choice of the collision parameters in our idealized simulations.

\end{document}